\documentclass[pra,twocolumn,amsmath,amssymb,superscriptaddress,showpacs]{revtex4}
\usepackage{graphics}
\usepackage{epsfig}
\usepackage{color}
\usepackage{pdfpages}

\usepackage{ucs}
\usepackage[utf8x]{inputenc}

\newcommand{\comment}[1]{}

\newcommand{\bra}[1]{\ensuremath{\left\langle{#1}\right\vert}}

\newcommand{\ket}[1]{\ensuremath{\left|{#1}\right\rangle}}

\def\be{\begin{equation}}
\def\ee{\end{equation}}
\def\eea{\end{eqnarray}}
\def\bea{\begin{eqnarray}}

\newcommand{\va}[1]{\ensuremath{(\Delta#1)^2}}

\newcommand{\ex}[1]{\ensuremath{\langle{#1}\rangle}}

\usepackage{bbm} 
\newcommand{\kommentar}[1]{}

\newcommand{\singlet}{\Psi^-}
\newcommand{\STheta}{S}
\newcommand{\CTheta}{C}

\newcommand{\qed}{\ensuremath{\hfill \Box}}

\newcommand{\indchain}{{\rm c}}							% index for general chain, single element
\newcommand{\indechain}{{\rm ce}}					% index for general chain, vector
\newcommand{\indchainV}{{\rm \ c}}					% index for equidistant chain, single element
\newcommand{\indfprod}{{\rm p}}
\newcommand{\indwn}{{\rm wn}}  							% index for white noise

\newcommand{\Jc}{J_x}
\newcommand{\JcB}{J_y}
\newcommand{\Jcconst}{J_z}
\newcommand{\jc}{j_x}
\newcommand{\jcB}{j_y}
\newcommand{\JJc}{X}

\newcommand{\jfield}{j_z}

\newcommand{\actualize}[2]{#2}

\begin{document}
\title{Macroscopic singlet states for gradient magnetometry}
\date{\today}
\begin{abstract}
We present a method for measuring magnetic field gradients with macroscopic singlet states 
realized with ensembles of spin-$j$ particles.
While the singlet state is completely insensitive to homogeneous
magnetic fields, the variance of its collective spin components is highly sensitive to field
gradients. We compute the dynamics of this variance analytically for a chain of spins and also
for an ensemble of particles with a given density distribution. 
 We find an upper bound on how precisely the field gradient can be estimated from the measured data.
 Based on our calculations, differential magnetometry can be carried out with cold atomic ensembles using a multipartite singlet state obtained via spin squeezing. On the other hand, comparing the metrological properties of the
 experimentally prepared state to that of the ideal singlet
 can be used as further evidence that a singlet state has indeed been  created.

 \end{abstract}

\author{I\~nigo Urizar-Lanz}
\affiliation{Department of Theoretical Physics, University of the Basque Country UPV/EHU, P.O. Box 644, E-48080
Bilbao, Spain}
\author{Philipp Hyllus}
\affiliation{Department of Theoretical Physics, University of the Basque Country UPV/EHU, P.O. Box 644, E-48080
Bilbao, Spain}
\author{I\~nigo Luis Egusquiza}
\affiliation{Department of Theoretical Physics, University of the Basque Country UPV/EHU, P.O. Box 644, E-48080
Bilbao, Spain}
\author{Morgan W. Mitchell}
\affiliation{ICFO-Institut de Ciencies Fotoniques, Mediterranean Technology Park, E-08860 Castelldefels (Barcelona), Spain}
\affiliation{ICREA-Instituci\'{o} Catalana de Recerca i Estudis Avan\c{c}ats, E-08015 Barcelona, Spain}
\author{G\'eza T\'oth}
\email{toth@alumni.nd.edu}
\homepage[\\URL: ]{http://www.gtoth.eu}
\affiliation{Department of Theoretical Physics, University of the Basque Country UPV/EHU, P.O.  Box 644, E-48080
Bilbao, Spain}
\affiliation{IKERBASQUE, Basque Foundation for
Science,
E-48011 Bilbao, Spain}
\affiliation{Wigner Research Centre for Physics, Hungarian Academy of 
Sciences, P.O. Box 49, H-1525 Budapest,  Hungary}

%03.65.-w	Quantum mechanics [see also 03.67.?a Quantum information; 05.30.?d Quantum %statistical mechanics;
%31.30.J?	Relativistic and quantum electrodynamics (QED) effects in atoms, %molecules, and ions in atomic physics]
%
%42.50.Dv	Quantum state engineering and measurements (see also 03.65.Ud Entanglement and %quantum nonlocality,
%                        e.g., EPR paradox, Bells inequalities, GHZ states, etc.)
%
%42.50.-p		Quantum optics (for lasers, see 42.55.?f and 42.60.?v; see also 42.65.?k
% 			Nonlinear optics; 
%03.65.?w 	Quantum mechanics)
% 03.65.Fd	Algebraic methods (see also 02.20.-a Group theory)
% 42.50.Lc       Quantum fluctuations, quantum noise, and quantum jumps.
% 07.55.Ge      Magnetometers for magnetic field measurements
% 03.67.Bg	Entanglement production and manipulation
% 03.75.-b        Matter waves


\maketitle

%%%%%%%%%%%%%%%%%%%%%%%%%%%%%%%
\section {Introduction}
%%%%%%%%%%%%%%%%%%%%%%%%%%%%%%%

Realization of large coherent quantum systems is at the center of attention in quantum experiments with 
cold atoms \cite{Hald1999,Julsgaard2001} and trapped ions  \cite{Meyer2001}.
Besides creating large scale entanglement, it is also important to look for quantum information processing applications of the states created. 
Recently, a series of experiments has been carried out with cold atomic ensembles using spin squeezing \cite{Hammerer2010}. 
This approach makes it possible to entangle  
$10^6-10^{12}$ atoms with each other by making them interact with a light field and then measuring the light, realizing in this way a quantum nondemolition (QND) measurement of one of the collective spin components \cite{Kuzmich1998,Duan2000}. Spin squeezed states are useful for 
continuous variable quantum teleportation \cite{Sherson2006} and magnetometry 
\cite{Eckert2006,Wasilewski2010,Wolfgramm2010,Sewell2011,ShahPRL2010,HorromPRA2012}.
In these experiments the atomic ensembles were almost completely  polarized, which makes it possible to map the quantum state of 
these ensembles to the state of bosonic modes \cite{Giedke2002} and model with few variables even realistic dynamics including noise \cite{Madsen2004,Hammerer2004,deEchaniz2005,Koschorreck2009}. 

A basic scheme for magnetometry with an almost completely polarized spin squeezed state works as follows. The total spin of the ensemble is rotated by a magnetic field perpendicular to it. The larger the field, the larger the rotation, which allows one to obtain the field strength by measuring a spin component perpendicular to the mean spin. So far, it looks as if the mean spin behaves like a clock arm and its position will tell us the value of the magnetic field exactly. However, at this point one has to remember that we have an ensemble of particles governed by quantum mechanics, and the uncertainty of the spin component perpendicular to the mean spin is never zero. Spin squeezing \cite{Kitagawa1993,Wineland1994,Sorensen2001,Ma2011} can decrease the uncertainty of one of the perpendicular components and this can be used to increase the precision of the magnetometry 
\cite{Kuzmich1998}.

Often the interesting quantity is not the absolute strength of the magnetic field, but its gradient and the effect of the homogeneous field must be 
suppressed. For example, the magnetic field of Earth must be suppressed when the much smaller magnetic field around an electric device or magnetic structure is measured
\cite{Keenan2012}. The field gradient can be determined by differential magnetometry, which can be carried out when two completely polarized atomic ensembles are used.
In fact, the same light beam can pass through the two atomic ensembles, which can be used both for simultaneous spin squeezing of the two atomic ensembles and to carry out the differential
measurement 
 \cite{Eckert2006, Wasilewski2010}. In general, singlet states of two large spins also offer the possibility for differential magnetometry \cite{Cable2010}.
 
 \begin{figure}%[h!]
\centering
           \includegraphics[width=0.45\textwidth]{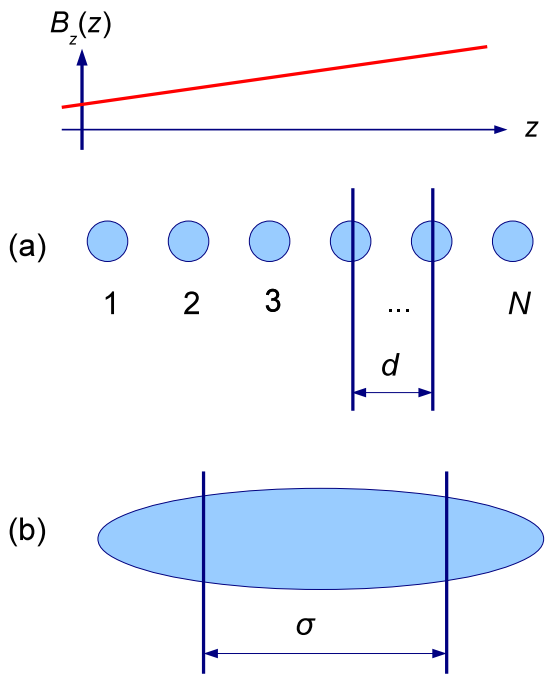}
\caption{(Color online) Multi-qubit singlet in a field gradient. (a) Equidistant chain of $N$ particles. (b) Atomic ensemble with a Gaussian density profile. 
\comment{One may put $z_0$ in the figure, cf. Eq. (21).}
}  \label{fig_chain_and_Gaussian}
 \end{figure}

It has recently been shown that interesting quantum states can be obtained even in unpolarized ensembles. In particular, if the uncertainties of the three
collective angular momentum components are squeezed one after the other, then a multiparticle singlet state 
can be obtained \cite{Toth2010,Behbood2012}. (For other approaches creating singlet states of cold atoms, see Ref.~\cite{Yao2011,Stevnson2011,Meineke2012}.)
Singlets, as ground states of antiferromagnetic Heisenberg spin systems, 
have attracted considerable attention \cite{Singlet1,Singlet2,Singlet3,Singlet4,Singlet5}.
Such states are invariant under the action of homogeneous magnetic fields. On the other hand, a magnetic field gradient rotates the spins at different locations differently, which leads to the gradual destruction of the singlet state. During this process, the variance of the collective spin components is increasing and this fact can be used to measure the field gradient. The advantage of this method is that only a single ensemble is used for differential magnetometry, rather than two ensembles, which leads to a better resolution and also eases the experimental requirements of the method. The basic scheme for differential magnetometry with singlets is depicted in Fig.~\ref{fig_chain_and_Gaussian}. (Other methods for measuring the field gradient can be found, for example, in Refs.~\cite{Koschorreckt2011, Zhou2010,BM13,Vengalattore2007,Wildermuth2006}.)

Besides demonstrating the usefulness of multiparticle singlets for metrology, our findings are interesting also to make singlets ``visible" in an experiment. An insensitivity to homogeneous fields and a growing angular momentum variance due to a field gradient can be strong evidence that a singlet state has indeed been created. If we change the sign of the gradient, then the variances should start to decrease. Such a gradient echo experiment can be another way to analyze singlets. The accuracy of the magnetic field gradient measurement achieved in an experiment can also be compared to our calculations, and such a comparison could be used as further evidence that a singlet has been created.

Finally, our calculations are interesting from the theoretical point of view  since we
succeed in modeling the quantum dynamics of large atomic ensembles 
analytically. This is a surprise as quantum systems with millions of particles are typically difficult to model.

While we mainly discuss spin-$\frac{1}{2}$ particles,
the spin squeezing procedure creating singlets and the differential metrology presented in our 
paper work also for spin-$j$ particles for $j>\frac{1}{2}.$
This is very important as looking for applications of ensembles of spin-$j$ particles without restricting the dynamics to the spin-$\frac{1}{2}$ subspace is
at the center of attention recently from the point of view of experiments and experimental proposals 
\cite{Sewell2011,Dubost2011,Hamley2012,LS12,Klimov2011,Toth2010,Kurucz2011,Sau2011,Mias2008,StamperKurn2012}, 
and also from the point of view of spin squeezing entanglement criteria \cite{Sorensen2001b,Duan2002,Mustecaplioglu2002,Toth2004,Wiesniak2005,Vitagliano2011,He2011}.

Our paper is organized as follows. In Sec.~\ref{sec:singlets}, we describe the multiparticle singlet state of spin-$\frac{1}{2}$ particles and analyze its properties.
In Sec.~\ref{sec:grad_mag}, we calculate the dynamics of the variance of collective angular momentum components for such states under a magnetic field gradient for a singlet realized with a spin chain shown in Fig.~\ref{fig_chain_and_Gaussian}(a). We also compute bounds on the precision of the gradient measurements.  In Sec.~\ref{sec:cont_dens}, we present calculations for an atomic ensemble with a given density profile depicted in Fig.~\ref{fig_chain_and_Gaussian}(b).
In Sec.~\ref{sec:further}, we 
consider the case of the singlet state mixed with noise and 
present results for particles with a spin larger than $\frac{1}{2}.$
Finally, we conclude the article in Sec.~\ref{sec:disc}.

%%%%%%%%%%%%%%%%%%%%%%%%%%%%%%%
\section {Multi-particle singlet states}
\label{sec:singlets}
%%%%%%%%%%%%%%%%%%%%%%%%%%%%%%%

In this section, we present an efficient description of multiparticle singlet states
of $N$ spin-$\frac{1}{2}$ particles.

Pure multiparticle singlet states are eigenstates of $J_l$ with a $0$ eigenvalue for $l=x,y,z.$
Here, the collective operators are defined as 
\begin{equation}
J_l=\sum_{n=1}^{N} j_l^{(n)}, \label{coll}
\end{equation}
where $j_l=\tfrac{\hbar}{2}\sigma_l$ with the Pauli spin matrices $\sigma_l$ for $l=x,y,z.$
Due to this, pure multiparticle singlet states are invariant under the unitary transformations
\begin{equation}
U_{\vec{n}}(\theta)=\exp\left(-i\frac{J_{\vec{n}}}{\hbar}\theta\right), \label{UUUU}
\end{equation}
where the angular momentum component along the $\vec{n}$ direction is
\begin{equation}
J_{\vec{n}}=n_x J_x+n_y J_y+n_z J_z.
\end{equation}
Such unitary transformations can be written as
$U_{\vec{n}}(\theta)=u^{\otimes N},$ where $u=\exp\left(-\frac{i}{\hbar}\sum_l n_l j_l\theta\right).$

Mixed multiparticle singlet states are mixtures of pure multiparticle singlets. Hence,
multiparticle singlets give zero for the expectation values of all moments of all collective angular momentum components
\begin{equation}
\ex{J_l^m}=0, \label{eqsinglet}
\end{equation}
where $l=x,y,z$ and $m=1,2,...,N.$ 
Mixed multipartite singlets are also invariant under the transformations of the type given by Eq.~(\ref{UUUU}).

In summary, singlet states are the states within the zero subspace of the Hamiltonian
\begin{equation}
H_{\rm s}=\kappa (J_x^2+J_y^2+J_z^2),
\end{equation}
where $\kappa>0$ is a constant.
The dimension of this space is growing rapidly with $N$ \cite{Cirac1999}. 
We have to identify the singlet created in the spin squeezing procedure in this space.

\subsection{Determining the singlet obtained in spin squeezing experiments}

In this section, we determine the multiparticle singlet created by spin squeezing
procedures. Due to symmetries of the setup,
the state created is permutationally invariant. There are very many multiparticle singlets. 
We will now show that, on the other hand, there is a unique  permutationally invariant singlet. 

Permutational invariance  means that the quantum state $\varrho$ equals its permutationally invariant part
\bea
(\varrho)_{\mathrm{PI}}=\frac{1}{N!}\sum_{k=1}^{N!} \Pi_k \varrho \Pi_k^\dagger, \label{PI}
\eea
where $ \Pi_k$ is a permutation operator and the summation is over all permutations.
The singlet state realized by the squeezing procedure in an atomic ensemble is permutationally invariant for the following reasons.
First, it is created starting from the completely
mixed state, which is permutationally invariant. Second, the
measurement-feedback procedure to squeeze the collective
variables involves only collective, not individual
variables. Hence, the dynamics is 
completely symmetric under exchange of particles \cite{Toth2010}.

Hence, we can state the following.

{\bf Observation 1.} 
{\it
For a given even number of particles $N$, there is a unique permutationally invariant singlet state.
It can be expressed as
\bea 
\varrho_{\rm s} =\lim_{T\rightarrow 0} \frac{e^{-\frac{H_{\rm s}}{T}}}{{\rm Tr }(e^{-\frac{H_{\rm s}}{T}})}  \label{singlet_J0}
\eea
and
\bea 
\varrho_{\rm s} =( \ket{\singlet} \bra{\singlet} 
\otimes \cdots \otimes \ket{\singlet} \bra{\singlet})_{\rm PI}, \label{singlet_pairs}
\eea
where the operation $(X)_{\rm PI}$ is defined in Eq.~(\ref{PI}) and the two-particle singlet is
\bea
\ket{\singlet}=\frac{1}{\sqrt{2}}\left(\ket{+\tfrac{1}{2},-\tfrac{1}{2}}_z-\ket{-\tfrac{1}{2},+\tfrac{1}{2}}_z\right),
\eea
where $\ket{\pm\tfrac{1}{2}}_z$ are the eigenstates of $j_z.$} \\
{\it Proof.} First we will show that there is a unique permutationally invariant singlet state. 
Such states have the following properties:
(i) they are permutationally invariant and (ii) they are the eigenstates of $J_l$ for $l=x,y,z$ with eigenvalues $0.$ 
All permutationally invariant multiparticle states
are uniquely characterized by the expectation values $\langle A^{\otimes (N-n)} \otimes \openone^{\otimes n} \rangle$, where $n=0,1,2,...,N-1$ and $A$ is a
traceless single-particle operator \cite{TothPRL}.
Moreover, as discussed before, all states for which $J_l=0$ for $l=x,y,z$ are
 invariant under the transformations of the type $u^{\otimes N},$ where $u$ are unitary matrices acting on a single spin. Since any traceless $A$ can be obtained from $\sigma_z$ by unitaries,
such a state can be uniquely characterized by the $N$ expectation values
$\langle \sigma_z^{\otimes (N-n)} \otimes \openone^{\otimes n}\rangle$, where $n=0,1,2,...,N-1$. Knowing these expectation values is
the same as knowing the expectation values of the powers $J_z^{n}$, for $n=1,2,...,N$.
However, these expectation values are zero for all singlets as can be seen in Eq.~(\ref{eqsinglet}). 
Thus, there is a single permutationally invariant singlet state, and Eqs.~(\ref{singlet_J0}) and (\ref{singlet_pairs}) are indeed equal \cite{JC12}. $\qed$

Let us interpret first  the Eq.~(\ref{singlet_J0})  formula. It denotes a state that is a completely mixed state within the $\sum_l \ex{J_l^2}=0$ subspace. It can also be written as
\begin{equation}
 \varrho_{\rm s} = \frac{1}{d_0} \sum_{\alpha =1}^{d_0} \ket{0,0,\alpha} \bra{0,0,\alpha},
\end{equation}
where $\ket{j,j_z,\alpha}$ denotes a state for which $\sum_l J_l^2\ket{j,j_z,\alpha}=j(j+1)\ket{j,j_z,\alpha},$
$J_z\ket{j,j_z,\alpha}=j_z\ket{j,j_z,\alpha},$  $\alpha$ is used to label the degenerate eigenstates, and $d_0$
is the degeneracy of the $j=j_z=0$ eigenstate \cite{Cirac1999}.

An alternative expression for the permutationally invariant singlet is given in Eq.~(\ref{singlet_pairs}). It shows that the multiparticle singlet is an equal mixture of all tensor products of two-particle singlets \cite{Note2}.
One can even find that for an even $N,$ the number of different permutations of such a singlet chain is  \cite{Note}
\begin{equation}
f(N)=(N-1)!!=(N-1)(N-3)(N-5)\cdot...
\end{equation}
 This fact has very important consequences for modeling quantum systems in such a state.
While storing the density matrix for a general quantum state of many particles is impossible, storing a representation of a product
state or a state that is a product of few-particle units can be done efficiently. In the next sections we will explain how to compute quantum dynamics starting from the permutationally invariant multiparticle singlet state.

\subsection{Calculating the reduced states of the singlet}
\label{sec:rho_red}

In this section, we will calculate $\varrho_{\rm 1}^{\rm red},$
$\varrho_{\rm 12}^{\rm red},$ and $\varrho_{\rm 1234}^{\rm red},$ 
which are respectively 
the reduced one-particle, two-particle, and four-particle density matrices of the singlet state $\varrho_{\rm s}.$ 
Later, this will be needed when computing the time evolution of certain operators
for the singlet state.

We have to start from the decomposition given by Eq.~(\ref{singlet_pairs}). From that, we obtain the form
\bea
\rho_{1234}^{\rm red}&=& \bigg(\alpha \frac{\openone}{16}+\beta\ket{\Psi_{12}^-}\bra{\Psi_{12}^-}\otimes \ket{\Psi_{34}^-}\bra{\Psi_{34}^-}\nonumber\\
&&+\gamma\ket{\Psi_{12}^-}\bra{\Psi_{12}^-}\otimes \frac{\openone}{4}+\text{permutations}\bigg), \label{rho1234}
\eea
where the second term has altogether three different permutations 
while the third term has six different permutations [including those appearing in Eq.~(\ref{rho1234})].
The term multiplied by $\beta$ is obtained when the particles $1$ to $4$ are in a product of two singlet states. The number of times that this special order is reached is given by the number of ways to distribute the remaining $(N-4)$ particles in $(N-4)/2$ pairs, that is, $(N-4-1)!!.$ This has to be divided by the number of {\it all} distributions, that is, $(N-1)!!$. Hence we obtain
\bea
\beta=\frac{(N-4-1)!!}{(N-1)!!}=\frac{1}{(N-1)(N-3)}.
\eea
The term multiplied by $\gamma$ is obtained when the particles $1$ and $2$ are in a singlet state, but particles $3$ and $4$ are not. As above,
the number of times  the remaining $N-2$ particles can be distributed in pairs is $(N-2-1)!!.$ However, we have to subtract here the number of distributions where particles $3$ and $4$ are in a singlet state as well, that is, $(N-4-1)!!$ as shown above. Again, we have to divide this by the 
total number of distributions, $(N-1)!!$, arriving at
\bea
\gamma&=&\frac{(N-2-1)!!}{(N-1)!!}-\frac{(N-4-1)!!}{(N-1)!!}\nonumber\\
&=&\frac{1}{(N-1)}-\frac{1}{(N-1)(N-3)}.
\eea
Finally, for the coefficient of the completely mixed component we have
\bea
\alpha&=&1-3\beta-6\gamma.\nonumber \\
\eea
This occurs when all four particles are in a singlet state with 
particles outside this set.

The two-spin reduced density matrix can be obtained from Eq.~(\ref{rho1234}) by tracing out particles 3 and 4 as
\bea
\rho_{12}^{\rm red}={\rm Tr}_{34}  (\rho_{1234}^{\rm red})=p_{\rm s}  \ket{\Psi_{12}^-}\bra{\Psi_{12}^-}+ (1-p_{\rm s}) \frac{\openone}{4} ,\label{rho12}
\eea
where 
\bea
p_{\rm s}=\beta+\gamma=\frac{1}{N-1},
\eea
which can also be obtained from combinatorial calculations similar to the ones
carried out for $\beta$ as $(N-3)!!/(N-1)!!.$
Finally, this leads to the trivial single-spin reduced state
\be
	\rho_1^{\rm red}=\frac{\openone}{2}.
	\label{rho1}
\ee

%%%%%%%%%%%%%%%%%%%%%%%%%%%%%%%
\section {Gradient magnetometry with a multiparticle spin chain}
\label{sec:grad_mag}
%%%%%%%%%%%%%%%%%%%%%%%%%%%%%%%

In this section, we consider $N$ spin-$\frac{1}{2}$ particles
in a permutationally invariant singlet state of Eq.~(\ref{singlet_pairs}),
where the particles are confined to a one-dimensional array. 
This could be prepared, for instance, with a Bose-Einstein 
condensate in an optical lattice driven to the so-called
Mott-insulator state \cite{GreinerNat02,MorschRMP06}.

We will calculate the effect of the magnetic field gradient on the singlet.
In particular, we will calculate how the variance of the collective
angular momentum components increases with the application of the field
gradient. We will also calculate how precisely the field gradient
can be estimated from the measured data.

In our calculations, 
we consider a one-dimensional array along the $z$-direction,
such that the positions of the particles are given by
\bea
(x_n,y_n,z_n)=(0,0,z_n^\indchain),
\eea
where $n=1,2,..,N$. 
%We assume that $z_n^0\le z_{n'}^0$ if $n < n'$ without loss of generality. 
A particular example is the equidistant chain
where 
\be 
	z_n^\indechain=(n-1)\cdot d+z_0, 
	\label{eq:equidist_chain}
\ee	
$d$ is the distance
between the particles and $z_0$ is an offset.
This situation is depicted in Fig.~\ref{fig_chain_and_Gaussian}.
We collect the positions on the $z$-axis
in the vector $\vec z_N^\indchainV$.

We can write the field at the atoms,  situated along $x=y=0$, as
\begin{equation}
\mathbf{B}(0,0,z)=\mathbf{B}_0 +z \mathbf{B}_1 + O(z^2),
\end{equation}
where we will neglect the terms of order two or higher.
We will consider 
$\mathbf{B}_0=B_0 \cdot (0,0,1)$ and
$\mathbf{B}_1=B_1 \cdot (0,0,1).$
For this configuration, due to the Maxwell equations, for the case of no currents or changing electric fields, we have
\begin{eqnarray}
{\rm div} \; \mathbf{B}&=&0, \nonumber \\
{\rm curl} \; \mathbf{B}&=&\mathbf{0}.
\end{eqnarray}
This implies $\sum_{l=x,y,z} \partial B_l /\partial l=0$ and $ \partial B_l /\partial m- \partial B_m /\partial l=0$ for $l\ne m.$
Thus, the spatial derivatives of the field components are not independent of each other. 
However, in the case of a linear chain  only the derivative along the chain has an influence on the quantum 
dynamics of the atoms. A similar statement holds for a quasi one-dimensonal atomic ensemble, which is typically the case if we consider an elongated trap.

The Hamiltonian corresponding to the effect of a homogeneous magnetic field in the $z$-direction is
\begin{equation}
H_{\hat z}=\gamma B_0 \sum_{n=1}^{N} \jfield^{(n)},
\end{equation}
where $\gamma$ is the gyromagnetic ratio. 
{It gives rise to a time evolution given in Eq.~(\ref{UUUU}) with $\vec n=\hat z$,
where $\hat z$ is the unit vector pointing in the $z$-direction.}
As we have discussed before, multiparticle singlets are invariant under the transformations of the type given by Eq.~(\ref{UUUU}).
On the other hand, the singlet is not invariant under the quantum dynamics 
generated by a magnetic field {\it gradient} $B_1$ described by the Hamiltonian
\begin{equation}
H_{\rm G}= \gamma B_{1} \sum_{n=1}^{N} z_n^\indchain \jfield^{(n)},\label{HG}
\end{equation}
where $z_n^\indchain$ is the position on the $z$-axis of the spin $n$ 
and $B_{1}$ is the field gradient along the $z$ direction. 
{Introducing a characteristic length $L$,
the Hamiltonian (\ref{HG}) can be rewritten as
\be
H_{\rm G}= 
\omega_L\sum_{n=1}^{N} \Big(\frac{z_n^\indchain}{L}\Big)\jfield^{(n)}, \label{HG_chain}
\ee
where $\omega_L=\gamma B_1 L$.}

{For instance, one may choose $L=d$ in the case of the equidistant chain described by 
Eq.~(\ref{eq:equidist_chain}).
Introducing the normalized Hamiltonian $H_{\rm G}'= \frac{H_{\rm G}}{\omega_L}$
and
\begin{equation}
\Theta=\omega_L t,
\end{equation}
the time evolution operator becomes
\be
	U_{\rm G}(\Theta)=\exp[-i  \frac{H_{\rm G}'}{\hbar} \Theta].
	\label{eq:Ugradient}
\ee
This formalism expresses the fact that our setup measures the field gradient times the time.
}

In order to use the singlet for differential magnetometry, we need to 
{
find an observable that changes with $\Theta$.
We investigate powers of collective operators $J_l^m$ since these
are relatively easy to measure experimentally.
It is clear that $l=z$ is not a good choice because 
$[J_z,H_{\rm G}']=0$, so this observable does not change with $\Theta$.
Therefore, we start by investigating whether the first or 
second moment of $J_x$ might be a suitable candidate.
}

%%%%%%%%%%%%%%%%%%%%%%%%%%%%%%%
\subsection{Calculating $\ex{\Jc}$ and $\ex{\Jc^2}$}
\label{sec:Jx2chain}
%%%%%%%%%%%%%%%%%%%%%%%%%%%%%%%

We compute the dynamics of $\ex{J_x}$ and $\ex{\Jc^2}$ starting from 
a multipartite singlet taking advantage of the fact that it is the mixture of all the permutations of tensor products of two-particle singlets as can be seen in Eq.~(\ref{singlet_pairs}). 
We will work in the Heisenberg picture, thus all operators will be given as a function of $\Theta,$ while expectation values
will be computed for the initial state $\varrho_{\rm s}.$
Hence the time evolution of the operator $J_x$ is given with the time-dependent single-spin operators as
\bea
\Jc(\Theta)=\sum_n \jc^{(n)}(\Theta). \label{Jx}
\eea

\subsubsection{Chain of particles at arbitrary positions} 

In this part, we will carry out calculations for a chain of particles at arbitrary positions. 
In the next part, we will present results for the equidistant chain.

In the Heisenberg picture, the time dependence of an operator $A$ is given as
\begin{equation}
A(\Theta)=\exp \left( i\frac{H_{\rm G}'}{\hbar} \Theta \right)A\exp \left( -i\frac{H_{\rm G}'}{\hbar} \Theta \right).\label{Heisenberg}
\end{equation}
Hence,
for the time dependence of  the single-particle operator $j_z^{(n)}$ we obtain
\begin{equation}\label{eq:Xop}
 \jc^{(n)}(\Theta)=c_n  \jc^{(n)}-s_n  \jcB^{(n)} \equiv \JJc_{\Theta}^{(n)},
\end{equation}
where we introduced the notation 
{
\begin{equation}
c_n= \cos\Big(\frac{z_n^\indchain}{L} \Theta\Big) \text{ and }s_n=\sin\Big(\frac{z_n^\indchain}{L}\Theta\Big). \label{defsandc}
\end{equation}
The quantitites $c_n$ and $s_n$ are the cosine and sine of the phases picked up by the 
$n^{\rm th}$ particle. The expectation value of $J_x$ is therefore given by
\be
	\ex{J_x}_{\rm s}^{\vec z_N^\indchainV}(\Theta)=
	\sum_n \left(c_n\ex{j_x^{(n)}}_{\rho_1^{\rm red}}-s_n\ex{j_y^{(n)}}_{\rho_1^{\rm red}}\right)=0
\ee
because the single-particle reduced state of the singlet $\varrho_{\rm s}$
is the completely mixed state, cf. Eq.~(\ref{rho1}). Note 
that the reduced state $\rho_1^{\rm red}$
is the same for any $n$ since $\varrho_{\rm s}$ 
is permutationally invariant.
In analogy, it can be shown that $\ex{J_y}_{\rm s}^{\vec z_N^\indchainV}(\Theta)=0$.
Therefore, a measurement of $J_l$ is not suitable for estimating 
$\Theta$ for $l=x,y,z$. We continue by investigating whether a
measurement of $J_x^2$ is useful.
}

{We can write $J_x^2$ in the Heisenberg picture [cf. Eqs.~(\ref{Jx}) and ~(\ref{Heisenberg})]
as a sum over two variables.} Knowing that $(\JJc_{\Theta}^{(n)})^2=\frac{\openone}{4},$ we can write
the expectation value of the second moment of $\Jc$  as a sum
of a constant and two-body correlations as
\bea
\ex{\Jc^2}^{\vec z_N^\indchainV}_{\rm s}(\Theta)=\frac{N}{4}\hbar^2+\sum_{n_1 \neq n_2} \ex{\JJc_{\Theta}^{(n_1)}\JJc_{\Theta}^{(n_2)}}_{\varrho_{\rm s}}.\label{Jx2b}
\eea
{In the following, we will partly drop the index $\vec z_N^\indchainV$ and 
sometimes also $\Theta$ from $\ex{J_x^2}_{\rm s}^{\vec z_N^\indchainV}(\Theta)$
when there is no risk of confusion}.
Since the singlet state is permutationally invariant, the correlation term on the right-hand side of Eq.~(\ref{Jx2b}) can be rewritten with the two-particle reduced state of the singlet as 
\bea
&&\sum_{n_1 \neq n_2} \ex{\JJc_{\Theta}^{(n_1)}\JJc_{\Theta}^{(n_2)}}_{\varrho_{\rm s}}\nonumber\\
&&\;\;\;\;\;\;\;\;\;=\sum_{n \neq m} \ex{(c_n \jc^{(1)}-s_n \jcB^{(1)})(c_m\jc^{(2)}-s_m \jcB^{(2)})}_{\varrho_{12}^{\rm red}}.\nonumber\\\label{n1n2}
\eea
{In Sec.~\ref{sec:rho_red}, we obtained $\varrho_{12}^{\rm red},$ the reduced two-spin state of the singlet [cf. Eq.~(\ref{rho12})].}
Direct calculation shows that 
\bea
\ex{j_k\otimes j_l}_{\varrho_{12}^{\rm red}}=-\frac{\hbar^2}{4(N-1)}\delta_{kl}, \label{deltakl}
\eea
where $k,l=x,y,z$ and $\delta_{kl}$ is $1$ if the two indices are equal, otherwise it is $0.$
Substituting  Eq.~(\ref{deltakl}) into Eq.~(\ref{n1n2}), we obtain the sum of the two-body correlations as
\be
\sum_{n_1 \neq n_2} \ex{\JJc_{\Theta}^{(n_1)}\JJc_{\Theta}^{(n_2)}}_{\varrho_{\rm s}}=-\frac{\hbar^2}{4(N-1)}I_2, \label{cosmntheta2}
\ee
{where
\be
	I_2\equiv\sum_{n\neq m}(c_n c_m + s_n s_m), \label{eq:I2}
\ee
and therefore
\be
\ex{\Jc^2}^{\vec z_N^\indchainV}_{\rm s}(\Theta)=\frac{N\hbar^2}{4}
\Big[1-\frac{1}{N(N-1)}I_2\Big],\label{Jx2c}
\ee
Since this is a non-trivial function of $\Theta$, a measurement
of $J_x^2$ can be used to estimate the magnetic field gradient.}

Note that Eq.~(\ref{Jx2c}) contains a sum over two variables such that the two variables are not allowed to be equal. 
{For practical purposes, it is more useful to rewrite this with
independent sums that require less computational effort as
}
\be
I_2=\sum_{n,m} (c_nc_m+s_ns_m)-\sum_{n} (c_n^2+s_n^2)\equiv 
(\CTheta^2+\STheta^2-N),\label{cosmntheta2b}
\ee
where we define the sums
{
\bea
\CTheta&=&\sum_n \cos\Big(\frac{z_n^\indchain}{L}\Theta\Big), \nonumber\\
\STheta&=&\sum_n  \sin\Big(\frac{z_n^\indchain}{L}\Theta\Big), \label{CS}
\eea
and use that $c_n^2+s_n^2=1$.
}
A similar subtraction procedure can be used in a more complicated 
calculation for the fourth moment of an angular momentum component below
\cite{EPAPS}.
Inserting Eq.~(\ref{cosmntheta2b}) into Eq.~(\ref{Jx2c}), 
we can state the following.

{
{\bf Observation 2. }
{\it 
The time dependence of the expectation value of the second moment of $\Jc,$ starting from a singlet of a chain of 
$N$ particles with the $z$-coordinates $\vec z_N^\indchainV$, is given by
\bea
&& \ex{\Jc^2}^{\vec z_N^\indchainV}_{\rm s}(\Theta)=\frac{N\hbar^2}{4}
\Big\{1+\frac{1}{N(N-1)}[N-\CTheta^2-\STheta^2]\Big\}, 
\nonumber\\ \label{FormulaLarga}
\eea
where $C$ and $S$ are defined in Eq.~(\ref{CS}).
}
}

{
Due to the symmetries of the setup, the dynamics of 
the variance of the $y$ component of the angular momentum is the same as the dynamics of the variance of the
$x$ component,
\bea
\ex{\JcB^2}_{\rm s}(\Theta)=\ex{\Jc^2}_{\rm s}(\Theta),
\eea
while $\ex{\Jcconst^2}_{\rm s}(\Theta)=0$ as mentioned before.
}

{
From Sec.~\ref{sec:singlets} we know that the singlet at $t=0$ is invariant 
under the influence of any homogeneous magnetic fields. For $t>0$ this is not true any more. However, the singlet evolving under the influence of $H_{\rm G}$ given in Eq.~(\ref{HG_chain}) still remains invariant under the transformation 
$U_{\hat z}(\theta)=\exp(-i\frac{J_z}{\hbar}\theta)$ since $[H_{\rm G},J_z]=0$. Because of that $\ex{J_z^2}_{\rm s}(\Theta)=0$ for all $\Theta.$
}

{
%---------------------------------------------------------------------------------%
\subsubsection{Equidistant chain} 
For the particular example of the equidistant chain, 
it is convenient to rewrite $\ex{J_x^2}_{\rm s}$ using the identity
\be
	c_n c_m+s_n s_m= \cos\Big(\frac{z_n^\indchain-z_m^\indchain}{L}\Theta\Big), \label{ccss_cos}
\ee
which, from Eqs.~(\ref{Jx2c}) and (\ref{cosmntheta2b}),
and with the $z$-coordinates $z_n^\indechain$
given in Eq.~(\ref{eq:equidist_chain})
leads to
\be
\ex{\Jc^2}_{\rm s}(\Theta)\label{eq:Jx2_equidist_chain}
=\frac{N\hbar^2}{4}\Big\{1-\frac{1}{N(N-1)}\Big[\sum_{n,m}
\cos\big([n-m]\Theta\big)-N\Big]
\Big\}
\ee
with the choice $L=d$. 
}

{
The variance of $\Jc$ starts from zero as follows from the properties
of the singlet state. It can also be seen from Eq.~(\ref{eq:Jx2_equidist_chain})
because each of the $N^2$ terms in the sum is equal to 1 at $\Theta=0$.
Then the variance grows up to around the level of the completely 
mixed state (white noise) {$\frac{\openone}{2^N}$},
\bea
\ex{\Jc^2}_\indwn=\frac{N}{4}\hbar^2. \label{Jc_cm}
\eea
To be more precise, for even $N$
\bea
&& \ex{\Jc^2}_{\rm s}
(\Theta=\pi)=\frac{N\hbar^2}{4}\Big[1+\frac{1}{N-1}\Big], 
\label{FormulaLarga2}
\eea
which is very close to $\ex{\Jc^2}_\indwn$ for large $N.$
The reason for this is that
$\sum_{n,m}\cos\big([n-m]\pi\big)=0$ because when $N$ is even, 
the number of terms where $n-m$ is even 
(such that $\cos\big([n-m]\pi\big)=+1$)
is equal to the number of terms where 
$n-m$ is odd (such that $\cos\big([n-m]\pi\big)=-1$).
At $\Theta=2\pi,$ $\ex{\Jc^2}_{\rm s}$ returns to $0$ because
again $\cos\big([n-m]2\pi\big)=1$ for all $n$ and $m$ 
since $n-m$ is an integer.
Obviously, for the equidistant chain, $\ex{J_x^2}_{\rm s}(\Theta)$
is a periodic function with a period time
\bea
T=\frac{2\pi}{\omega_{L=d}}=\frac{2\pi}{\gamma B_1 d}.
\eea
If there is not a complete revival then the particles are not arranged in an array such that the interparticle distance is
uniform over the chain.
This phenomenon can be used to characterize chains of atoms from the point of view
of the uniformity of the distribution of the atoms.
}

\begin{figure}%[h!]
\centering
            \includegraphics[width=0.45\textwidth]{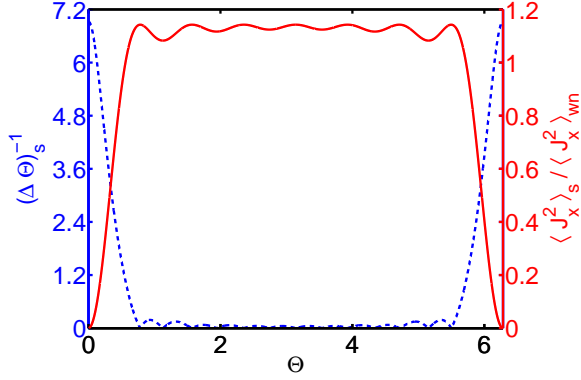}
            \vskip-0.5cm
\caption{(Color online) The dynamics of $\ex{\Jc^2}_{\rm s}/\ex{\Jc^2}_\indwn$ 
(solid line) and $(\Delta \Theta)_{\rm s}^{-1}$ (dashed line) as the function of $\Theta$ for 
an equidistant chain of $N=8$ particles. 
The angular momentum variance for the white noise, $\ex{\Jc^2}_\indwn,$  is defined in Eq.~(\ref{Jc_cm}).
}
\label{fig_Jx2_Fisher_N8}
 \end{figure}

In Fig.~\ref{fig_Jx2_Fisher_N8}, we plot the dynamics of $\ex{\Jc^2}_{\rm s}(\Theta)$ for $N=8$ spin-$\frac{1}{2}$ particles. Note that the {increase} of $\ex{\Jc^2}_{\rm s}$ is not monotonic for $\Theta<\pi,$ but, rather oscillating. This is due to the fact that the atoms are arranged in a lattice. \comment{Plot also for $10^5$ particles? Or does one not see anything in this case?}
In the case of a continuous density distribution, there is no such 
oscillation, as will be shown in Sec.~\ref{sec:cont_dens}.

%%%%%%%%%%%%%%%%%%%%%%%%%%%%%%%
\subsection {Calculating the precision of estimating $\Theta$}
\label{sec:CalcPrec}
%%%%%%%%%%%%%%%%%%%%%%%%%%%%%%%

{
From the calculation of $\ex{\Jc^2}_{\rm s}(\Theta)$ it is clear that 
a measurement of $\Jc^2$ gives information about $\Theta;$ hence
it can be used to estimate an unknown value of $\Theta$.
In this section, we will calculate the precision 
$\Delta\Theta$ of the estimation, for the singlet state realized 
with a chain of spins with the $z$-coordinates $\vec z_N^\indchainV$, 
undergoing a quantum dynamics due to a magnetic field gradient.
}

The finite precision in reconstructing $\Theta$ comes from the fact that $\Jc^2$ can only be measured 
with some uncertainty. The precision of the reconstruction of $\Theta$ based on measuring $\Jc^2$ 
can be obtained as
\begin{equation}
(\Delta \Theta)_{\rm s}^{2} =  
\frac{\va{\Jc^2}_{\rm s}}{|\partial_{\Theta}\ex{\Jc^2}_{\rm s}|^2} , \label{DeltaTheta}
\end{equation}
where $(\Delta A)^2=\ex{A^2}-\ex{A}^2$ for an observable $A$.
In order to calculate the expression (\ref{DeltaTheta}), we need to know $\ex{\Jc^2}_{\rm s}(\Theta),$ which we have just obtained, and 
$\ex{\Jc^4}_{\rm s}(\Theta),$ 
which we are going to calculate now.

Let us consider again the one dimensional chain of particles.
For this, the Hamiltonian is given in Eq.~(\ref{HG_chain}).
As in Sec.~\ref{sec:Jx2chain}, we will work again in the Heisenberg picture. 
The time evolution of the operator $\Jc^4$ is
\bea
\Jc^4(\Theta)=\bigg(\sum_n \jc^{(n)}(\Theta)\bigg)^4.\label{Jz4}
\eea
The expectation value of this 
can be rewritten as a sum over four variables as
\be
\ex{\Jc^4}_{\rm s}=\sum_{n_1,n_2,n_3,n_4} \ex{\JJc_{\Theta}^{(n_1)}\JJc_{\Theta}^{(n_2)}\JJc_{\Theta}^{(n_3)}\JJc_{\Theta}^{(n_4)}}_{\varrho_{\rm s}}. \label{sum4}
\ee
{Again, we leave out the indices $\vec z_N^\indchainV$ and $\Theta$
for simplicity at the moment.}
We can rewrite Eq.~(\ref{sum4}) with sums in which the variables of the summation are not allowed to be equal as
\bea
\ex{\Jc^4}_{\rm s} & = & 
\sum_{n_1} \ex{(\JJc_{\Theta}^{(n_1)})^4}_{\varrho_{\rm s}} \nonumber \\
&&+ 3 \sum_{n_1 \neq n_2} \ex{(\JJc_{\Theta}^{(n_1)})^2(\JJc_{\Theta}^{(n_2)})^2}_{\varrho_{\rm s}}  \nonumber \\
&&+ 4\sum_{n_1\neq n_2}\ex{(\JJc_{\Theta}^{(n_1)})^3(\JJc_{\Theta}^{(n_2)})} _{\varrho_{\rm s}}\nonumber \\
&&+ 6 \sum_{\neq (n_1,n_2,n_3)} \ex{(\JJc_{\Theta}^{(n_1)})^2(\JJc_{\Theta}^{(n_2)})(\JJc_{\Theta}^{(n_3)})}_{\varrho_{\rm s}} \nonumber \\
&&+ \sum_{ \neq (n_1,n_2,n_3,n_4)} \ex{\JJc_{\Theta}^{(n_1)}\JJc_{\Theta}^{(n_2)}\JJc_{\Theta}^{(n_3)}\JJc_{\Theta}^{(n_4)}}_{\varrho_{\rm s}},  \nonumber \\ \label{sum4b}\eea
where $\sum_{\neq (i,j,k)}$ denotes summation over the indices $j,k,l$ such that none of them is equal to another one.
Based on Eq.~(\ref{sum4b}) and by 
making use of the fact that $(\JJc_{\Theta}^{(n)})^2=\frac{\openone}{4}\hbar^2$, we arrive at
\bea
\ex{\Jc^4}_{\rm s} &=& \frac{\hbar^4}{16} \big[N+3N(N-1)\big] \nonumber\\
&&+\hbar^2\bigg[1+\frac{3(N-2)}{2}\bigg] \sum_{n_1 \neq n_2}\ex{\JJc_{\Theta}^{(n_1)}\JJc_{\Theta}^{(n_2)}}_{\varrho_{\rm s}}\nonumber\\
&&+\sum_{\neq (n_1,n_2,n_3,n_4)} \ex{\JJc_{\Theta}^{(n_1)}\JJc_{\Theta}^{(n_2)}\JJc_{\Theta}^{(n_3)}\JJc_{\Theta}^{(n_4)}}_{\varrho_{\rm s}}.\nonumber \\ \label{Jx4}
\eea
The expectation value of the sum of the two-body correlations is given in Eq.~(\ref{cosmntheta2}). 
Next, we calculate the expectation values for the four-body correlations.
For that, we use the reduced four-particle density matrix presented {in Eq.~(\ref{rho1234}).}
For the reduced four-particle matrix we obtain
\bea
\ex{\jc^{(1)}\jc^{(2)}\jc^{(3)}\jc^{(4)}}_{\varrho_{1234}^{\rm red}}=\frac{\hbar^4}{16}\frac{3}{(N-1)(N-3)}, \nonumber \\
\ex{\jcB^{(1)}\jcB^{(2)}\jcB^{(3)}\jcB^{(4)}}_{\varrho_{1234}^{\rm red}}=\frac{\hbar^4}{16}\frac{3}{(N-1)(N-3)}, \nonumber \\
\eea
and
\bea
\ex{\jc^{(1)}\jc^{(2)}\jcB^{(3)}\jcB^{(4)}}_{\varrho_{1234}^{\rm red}}=\frac{\hbar^4}{16}\frac{1}{(N-1)(N-3)}.
\eea
The singlet state is invariant 
{under $U^{\otimes N}$ for any local unitary 
$U$. 
%In particular, this is true for $U=\sigma_l$ for $l=x,y$.
This leads} to  
$\ex{\jc^{(1)}\jcB^{(2)}\jcB^{(3)}\jcB^{(4)}}_{\varrho_{1234}^{\rm red}}=\ex{\jc^{(1)}\jc^{(2)}\jc^{(3)}\jcB^{(4)}}_{\varrho_{1234}^{\rm red}}=0$
\cite{Note_SU2invariance}.
Since the singlet is permutationally invariant, all the expectation values
with four-body correlations with  $\jc$ and  $\jcB$ follow. 
{Taking everything into account, we arrive at the expression
\bea
\ex{\Jc^4}_{\rm s} &=& \frac{\hbar^4}{16}\bigg\{ 3N^2-2N-\frac{6N-8}{N-1}I_2 \nonumber \\ 
&& \qquad+ \frac{3}{(N-1)(N-3)}I_4 \bigg\}, \label{MomAng}
\eea
where we defined
\be
I_4\equiv \sum_{\neq(k,l,m,n)} [ c_kc_lc_nc_m 
 +s_ks_ls_ns_m+2c_kc_ls_ns_m]. \label{eq:I4}
\ee
The term $I_4$ involves
a sum over four variables, which implies a large computational 
effort for large systems.
Similarly to what has been done with $I_2$ in Eq.~(\ref{cosmntheta2b}), we can
write $I_4$ as 
}
\be
I_4= \sum_{k,l,n,m}[c_kc_lc_nc_m 
+ s_ks_ls_ns_m+2c_kc_ls_ns_m]-P.\label{defS}
\ee
In $P$ we will write all the terms that appear in $\sum_{k,l,n,m}$ but do not appear in $\sum_{\neq(k,l,m,n)}$.
Let us study the first term of the right-hand side of Eq.~(\ref{defS}). We can rewrite it as
\bea
\sum_{k,l,n,m}\big[c_kc_lc_nc_m+s_ks_ls_ns_m+2c_kc_ls_ns_m\big]\nonumber\\
=\CTheta^4+\STheta^4+2\CTheta^2\STheta^2,
\eea
where $\CTheta$ and $\STheta$ are defined in Eq.~(\ref{CS}). Next, we have to determine $P.$
After determining it, we have to eliminate the sums containing conditions that multiple indices are unequal in a similar manner
by replacing the sum with another one without such a condition and substracting the difference.
Finally, we arrive at an expression with single-index sums only.
Using these results, we obtain the following \cite{EPAPS}.

{
{\bf Observation 3. } 
{\it 
The time dependence of the expectation value of the fourth 
moment of $\Jc,$ starting from a singlet of a chain of 
$N$ particles with the $z$-coordinates $\vec z_N^\indchainV$, is given by
\bea
\ex{\Jc^4}_{\rm s}^{\vec z_N^\indchainV}(\Theta)
&=& \frac{\hbar^4}{16}\bigg\{3N^2-2N-\frac{6N-8}{N-1}I_2\label{jx4X}\\
&& \qquad+\frac{3}{(N-1)(N-3)}I_4\bigg\}\nonumber
\eea
where $I_2$ and $I_4$ are defined in Eqs.~(\ref{eq:I2})
and (\ref{eq:I4}) and can be rewritten as
\bea
I_2&=&\JJc_{1,0}^2+\JJc_{0,1}^2-N,\nonumber\\
I_4
&=&\JJc_{1,0}^4 +\JJc_{0,1}^4 +2\JJc_{1,0}^2\JJc_{0,1}^2 \nonumber \\  
&&-6\JJc_{4,0}-6\JJc_{0,4} - 12\JJc_{2,2} +3\JJc_{2,0}^2 + 3 \JJc_{0,2}^2\nonumber \\ 
&&+8\JJc_{3,0}\JJc_{1,0} + 8\JJc_{0,3}\JJc_{0,1}+4\JJc_{1,1}^2 +8\JJc_{2,1}\JJc_{0,1}\nonumber \\ 
&&+8\JJc_{1,2}\JJc_{1,0}+2\JJc_{2,0}\JJc_{0,2}-6\JJc_{2,0}\JJc_{1,0}^2 -6\JJc_{0,2}\JJc_{0,1}^2\nonumber \\ 
&&- 2\JJc_{2,0} \JJc_{0,1}^2 - 2\JJc_{0,2} \JJc_{1,0}^2-8\JJc_{1,1}\JJc_{1,0}\JJc_{0,1} ,\nonumber\\
\label{III444}
\eea
where
\bea
\JJc_{k,l}=\sum_{n=1}^N c_n^k s_n^l
\eea
with $c_n$ and $s_n$ as defined in Eq.~(\ref{defsandc}).
}
}

{
Alternatively, it is possible to write the term $I_4$ in a more compact form as \cite{EPAPS}
\bea
I_4&=& N\left\{2(N-3)-4 N(N-2)\left|\hat{f}_1(\alpha)\right|^2\right.\nonumber\\
&&\qquad\qquad\quad+N^3\left|\hat{f}_1(\alpha)\right|^4 + N\left|\hat{f}_1(2\alpha)\right|^2
\nonumber\\
&& \left.\qquad\qquad\quad-2N^2\textrm{Re}\big[\hat{f}_1^2(\alpha)\hat{f}_1(2 \alpha)^*\big]\right\},
\label{III444B}
\eea
where 
\begin{equation}
	\hat f_1(\alpha)=\frac{1}{N}\sum_k e^{i\alpha z_k^\indchain}\quad \mathrm{and}\quad
	\alpha=\frac{\Theta}{L},
\end{equation}
which is also easier to compute.
}

Observations 2 and 3 make it possible to calculate the precision of the estimation $\Delta \Theta$ based on Eq.~(\ref{DeltaTheta}). It is possible to calculate analytically the precision for $\Theta=0.$ For that, we determined the zeroth and first order terms of the Taylor expansion of Eq.~(\ref{DeltaTheta}) using the expansion of sine and cosine up to second order. Hence, we can state the following.

{\bf Observation 4. } 
{\it The maximal precision of estimating $\Delta \Theta$ for {an equidistant} 
chain of $N$ spin-$\frac{1}{2}$ particles is characterized by 
\bea
(\Delta \Theta)_{\rm s}^{-2}(\Theta=0)=\frac{N^2+N^3}{12}=\Big[N\frac{\sigma^2}{L^2}\Big]\frac{N}{N-1}, \label{DeltaTheta_chain}
\eea
where $\sigma=L\sqrt{(N^2-1)/12}$ is the standard deviation of the equidistant chain,
and $L=d$.
}

{
A more general result on the precision at $\Theta=0$ will be presented
in Observation 7 below.
}

In Fig.~\ref{fig_Jx2_Fisher_N8}, we plot the dynamics of $(\Delta \Theta)_{\rm s}^{-1}$ for $N=8$
for the equidistant chain. 
As can be seen, the precision is maximal at $\Theta=0.$ Then, it decreases, however, this decrease is 
not monotonic and the precision is oscillating. In particular, based on Eq.~(\ref{DeltaTheta}) one can see that the 
precision is zero when the tangent of $\ex{\Jc^2}_{\rm s}(\Theta)$ is horizontal, i.e.,  $\partial_\Theta \ex{\Jc^2}_{\rm s}(\Theta)=0.$
%\comment{Also for $10^5$?}

%%%%%%%%%%%%%%%%%%%%%%%%%%%%%%%
\section{Continuous density profile}
\label{sec:cont_dens}
%%%%%%%%%%%%%%%%%%%%%%%%%%%%%%%

In this section, we work out the formulas describing the case of a one-dimensional continuous density profile.
We present the dynamics of the second and fourth moments of the collective angular momentum components
for this case.

In the case of a spin chain, the particles were placed in 
{the fixed positions $\vec z_N^\indchainV$}.
Now, while we still consider the case when the particles are localized in certain positions,
the distribution of $N$ particles is given by a distribution function $f_N,$ where
\bea
f_N(z_1,z_2,...,z_N)  {\rm d}z_1  {\rm d}z_2 ...  {\rm d}z_N
{\equiv f_N(\vec z_N) {\rm d}\vec z_N}
\eea
is the probability that particle $1$ is between $z_1$ and $z_1+ {\rm d}z_1,$  particle $2$ is between $z_2$ and $z_2+ {\rm d}z_2,$  etc.
{Without loss of generality, $f_N$ can be considered
invariant under the permutation of any two particles.
As before, we compute the average $\ex{J_x^2}_{\rm s}$ 
in order to estimate the magnetic field
gradient and also $\ex{J_x^4}_{\rm s}$ in order to
estimate the uncertainty $\Delta\Theta$. For a general distribution
function $f_N(\vec z_N)$, they are given by
\be
	\ex{J_x^2}_{\rm s}^{f_N}(\Theta)=\int {\rm d}\vec z_N 
	f_N(\vec z_N)\ex{J_x^2}_{\rm s}^{\vec z_N}(\Theta),
\ee
with $\ex{J_x^2}_{\rm s}^{\vec z_N}(\Theta)$ from
Eq.~(\ref{FormulaLarga}), and 
\be
	\ex{J_x^4}_{\rm s}^{f_N}(\Theta)=\int {\rm d}\vec z_N 
	f_N(\vec z_N)\ex{J_x^4}_{\rm s}^{\vec z_N}(\Theta),
	\label{Jx4fff}
\ee
with $\ex{J_x^4}_{\rm s}^{\vec z_N}(\Theta)$ from
Eq.~(\ref{jx4X}). 
Here and in the following, we substitute $\vec z_N^\indchainV$
with $\vec z_N$ in the expressions for fixed particle positions.
}

{
In order to compute particular examples, we need some properties
of the reduced $M$-particle distribution functions $f_M$ ($M\le N)$. 
These can be obtained from $f_N$ as}
\be
f_M(\vec z_M) = \int {\rm d}z_{M+1}...  {\rm d}z_{N} f_{N}(\vec z_N),
\ee
{where we used the shorthand notation $\vec z_M=(z_1,z_2,...,z_M)^T$ as above.
Note that due to the invariance under permutations of $f_N$, it does not matter
which of the $N-M$ particles are integrated over. Further, 
$f_M$ is permutationally invariant as well.
}

%%%%%%%%%%%%%%%%%%%%%%%%%%%%
\subsection{Distributions with Dirac $\delta$ functions}
 %%%%%%%%%%%%%%%%%%%%%%%%%%%%

{
For example, the distribution function of the chain considered
in the previous section can be written as \cite{EPAPS}
\begin{equation}\label{eq:fNz0}
f_N^{\vec z_N^\indchainV}(\vec z_N)
=\frac{1}{N!}\sum_{\neq(k_1,k_2,\ldots,k_N)}\prod_{j=1}^N \delta\left(z_j-z_{{k_j}}^\indchain\right)\,.
\end{equation}
Its reduced distribution functions are given by
 \begin{equation}\label{eq:fM}
f_M^{\vec z_N^\indchainV}(\vec z_M)=\frac{(N-M)!}{N!}\sum_{\neq(k_1,k_2,\ldots,k_M)}\prod_{j=1}^M \delta\left(z_j-z_{{k_j}}^\indchain\right)\,.
\end{equation}
}

{
The permutational invariance of $f_M$ simplifies many calculations.
For instance, the average of $I_2$ from Eq.~(\ref{eq:I2}) for 
a distribution $f_N(\vec z_N)$ becomes
\bea
	\ex{I_2}^{f_N}&=&\int {\rm d}\vec z_N I_2(\vec z_N)\label{eq:I2av}\\
&=&\sum_{n \neq m} \int {\rm d}\vec z_N f_N(\vec z_N)[c_n c_m+s_n s_m] \nonumber\\
&=&\sum_{n \neq m} \int  {\rm d}z_{n}  {\rm d}z_{m} f_2(z_n,z_m)[c_n c_m +s_n s_m]\nonumber\\
&=&N(N-1) \int  {\rm d}\vec z_2 f_{2}(\vec z_2)[c_1 c_2+s_1 s_2]. \nonumber
\eea
In analogy, we obtain the average of $I_4$ from Eq.~(\ref{eq:I4}) for 
$f_N(\vec z_N)$,
\bea
\ex{I_4}^{f_N}&=&\int {\rm d}\vec z_N f_N(\vec z_N)I_4(\vec z_N)\label{eq:I4av}\\
&=&\frac{N!}{(N-4)!}\int{\rm d}\vec z_4 f_4(\vec z_4) \nonumber\\
&&\qquad \times[c_1 c_2 c_3 c_4+s_1 s_2 s_3 s_4+2c_1 c_2 s_3 s_4].
 \nonumber
\eea
With these, it is possible to recover all of our previous results for spin chains.
}

 %%%%%%%%%%%%%%%%%%%%%%%%%%%%
\subsection{Independently and smoothly distributed particles}
 %%%%%%%%%%%%%%%%%%%%%%%%%%%%
 
 {
In the following, we will make the assumption that 
the system is a gas of particles that are 
uncorrelated in space, which means that 
the distribution function can be written as a 
product of single-particle distribution functions,
\be
	f_N^\indfprod(\vec z_N)=\Pi_{n=1}^N f_1(z_n). \label{uncorr}
\ee
Further, we will assume that $f_1$ is a smooth function which
does not contain Dirac delta functions, {\it i.e.}, points
with infinitely high density.
}
 
{Let us compute the expectation values of $J_x^2$ and $J_x^4$ 
as we did in the previous section, for a product distribution 
function of the form given in Eq.~(\ref{uncorr}). Considering
Eqs.~(\ref{Jx2c}) and (\ref{MomAng}), it becomes clear that
we only need to compute $\ex{I_2}^{f_N}$ and $\ex{I_4}^{f_N}$.
Since $f_2^\indfprod(\vec z_2)=f_1(z_1)f_1(z_2)$ in the uncorrelated case, 
we obtain from Eq.~(\ref{eq:I2av}) that
\be
	\ex{I_2}^{f^\indfprod_N}=N(N-1)(\tilde C^2+\tilde S^2),
\ee
where
}
\bea  
\tilde{C}=\int  {\rm d}z_1  f_1(z_1) \cos\bigg(\frac{z_1}{L} \Theta\bigg)
\label{CT} 
\eea
and
\bea 
\tilde{S}=\int  {\rm d}z_1  f_1(z_1) \sin\bigg(\frac{z_1}{L} \Theta\bigg). \label{ST} \eea 
Here the averaging is over 
the density distribution $f_1(z_1).$
With these, we have all of the ingredients to determine the dynamics of 
$\ex{\Jc^2}_{\rm s}$ for the gas from Eq.~(\ref{Jx2c}).

{\bf Observation 5. } 
{\it 
For an ensemble of particles with a 
{product 
distribution function $f_N^\indfprod$ from Eq.~(\ref{uncorr}),
}
we obtain the following dynamics for the expectation value of 
the second moment of $\Jc$:
\begin{equation}
\ex{\Jc^2}^{f_N^\indfprod}_{\rm s}(\Theta)=\frac{N\hbar^2}{4}\Big[1-\tilde C^2-\tilde S^2\Big],\label{angularmomentum2b}
\end{equation}
where $\tilde{C}$ and $\tilde{S}$ are defined in  Eqs.~(\ref{CT}) and  (\ref{ST}). }

In analogy, we obtain from Eq.~(\ref{eq:I4av}) that 
\be 
	\ex{I_4}^{f^\indfprod_N}=\frac{N!}{(N-4)!}(\tilde C^2+\tilde S^2)^2.
\ee
Inserting $\ex{I_2}^{f^\indfprod_N}$ and $\ex{I_4}^{f^\indfprod_N}$
into Eq.~(\ref{MomAng}), we obtain the dynamics of $\ex{J_x^4}_{\rm s}$.

{\bf Observation 6. } 
{\it 
For an ensemble of particles with a
{product 
distribution function $f_N^\indfprod$ from Eq.~(\ref{uncorr}),
}
we obtain the following dynamics for the expectation value for the 
fourth moment of $\Jc$ 
\bea
\ex{\Jc^4}^{f_N^\indfprod}_{\rm s} &=& 
\frac{N\hbar^4}{16}\bigg\{
3N-2
-(6N-8)(\tilde{C}^2+\tilde{S}^2)\nonumber \\ 
&& \qquad\quad + 3(N-2)(\tilde{C}^2+\tilde{S}^2)^2 \bigg\}, \label{MomAng_d}
\eea
where $\tilde C$ and $\tilde S$ are defined in the Eqs.~(\ref{CT}) and (\ref{ST}).
}

Note that due to the product structure of $f_N^\indfprod$, 
this expression is much simpler than the expression for the chain given in Eq.~(\ref{jx4X}). These formulas make it possible already to calculate the precision of the phase estimation. 

\subsubsection{Gaussian density profile}
\label{SecGaussianDensityProf}

One of the most common density profiles is the Gaussian profile. 
As an example, we will now calculate the dynamics for this case explicitly.
The Gaussian density profile is given as
\begin{equation}
f_1^{\rm gauss}(z_1)=\frac{1}{\sqrt{2\pi\sigma ^2}} e^{-\frac{(z_1-z_0)^2}{2 \sigma^2}}, \label{GaussianDensity}
\end{equation}
where $z_0$ is the coordinate of the point with the highest density and 
$\sigma$ is the width of the profile.
Substituting it into Eq.~(\ref{angularmomentum2b}), we obtain
\bea
\tilde{C}^2+\tilde{S}^2= e^{-\frac{\sigma^2}{L^2}\Theta^2}  \label{CSGaussian}
\eea
and hence
\bea
\ex{\Jc^2}_{\rm s}^{f^\indfprod_N}(\Theta)= \frac{N\hbar^2}{4} \bigg( 1-e^{-\frac{\sigma^2}{L^2} \Theta^2} \bigg). \label{GaussianJ}
\eea
Note that the maximum of  $\ex{\Jc^2}_{\rm s}(\Theta)$ is exactly the value for white noise; 
cf. Eq.~(\ref{Jc_cm}). 
Thus, in this case there is no overshooting,
and the noise does not become larger than that of the white noise, as was the case with the {equidistant} chain in Eq.~(\ref{FormulaLarga}).
 
\begin{figure}%[h!]
\centering
           \includegraphics[width=0.5\textwidth]{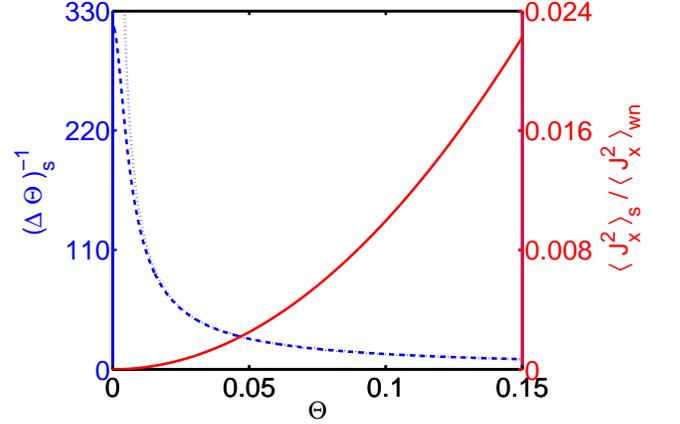}
           \vskip-0.5cm
\caption{(Color online) The dynamics of $\ex{\Jc^2}_{\rm s}/\ex{\Jc^2}_\indwn$ (solid line) and $(\Delta \Theta)_{\rm s}^{-1}$ (dashed line) as the function of $\Theta$ for $N=10^5$ particles {with a Gaussian density distribution
[Eq.~(\ref{GaussianDensity})]}
for $\sigma=L.$ We also present $(\Delta \Theta)_{\rm s}^{-1}$ calculated based on the Gaussianity assumption 
given by Eq.~(\ref{exJ4_Gaussian}) discussed in Sec.~\ref{Sec_Largej} (dotted line). The angular momentum variance for the white noise, $\ex{\Jc^2}_\indwn,$  is defined in Eq.~(\ref{Jc_cm}).
} \label{fig_Jx2_Fisher_Gauss}
 \end{figure}

In Fig.~\ref{fig_Jx2_Fisher_Gauss}, we plot the dynamics of $\ex{\Jc^2}_{\rm s}$ for $N=10^5$ spin-$\frac{1}{2}$ particles.
The density profile is a product distribution function of Gaussian profiles 
with $\sigma=L.$ \comment{Before: only 8 particles for the
equidistant chain! Should we harmonize that?}

\subsubsection{Analytic formula for the maximum precision for any particle distribution, including correlated particle distributions}

We will now give an analytic formula for the maximum precision at $\Theta=0$ for the case of any 
particle distribution.

{\bf Observation 7.} 
{\it 
The maximal precision of estimating $\Delta \Theta$ is characterized by
\bea\label{DeltaTheta_gas}
(\Delta \Theta)_{\rm s}^{-2}(\Theta=0)= N\Big[\frac{\sigma^{2}}{L^2}-\frac{\mathrm{cov}(z_{1},z_{2})}{L^2}\Big],
\eea
where }
\bea
\sigma^{2}&=&\int{\rm d}zf_{1}(z)(z-\langle z\rangle)^{2},\nonumber\\
\langle z\rangle&=&\int{\rm d}zf_{1}(z)z,\nonumber\\
\mathrm{cov}(z_{1},z_{2})&=&\int{\rm d}z_1 z_2 f_{2}(z_1,z_2)(z_1-\langle z_1\rangle)(z_2-\langle z_2\rangle).\nonumber\\
\eea
For the derivation, see the Supplemental Material \cite{EPAPS}. 
{For any product distribution
function $f_N^\indfprod$, $(\Delta \Theta)_{\rm s}^{-2}(\Theta=0)= N\frac{\sigma^{2}}{L^2}$ since
$\mathrm{cov}(z_{1},z_{2})=0$ in this case. Hence, the maximal
sensitivity is a simple function of $N$ and $\sigma$.
Correlations can increase or decrease the sensitivity compared to the 
uncorrelated case. For instance, the correlations present in the equidistant chain 
increase $(\Delta\Theta)_{\rm s}^{-2}(\Theta=0)$ by a factor $\frac{N}{N-1}$; cf. 
Eq.~(\ref{DeltaTheta_chain}). 
}

%---------------------------------------------------------------------------------%
\section{Further considerations}
\label{sec:further}

%%%%%%%%%%%%%%%%%%%%%%%%%%%%%%%
\subsection{The influence of noise}
%%%%%%%%%%%%%%%%%%%%%%%%%%%%%%%
\label{Sec_Noise}

So far, we considered only perfect singlet states.
In practice, the multipartite singlet state cannot be 
realized perfectly \cite{noise}. In this section, we will consider the 
case of starting from imperfect singlet states.
First we will discuss the case, when we measure $\langle J_x^2\rangle,$
as before. Then, we will show that a higher accuracy can be achieved in the noisy
case, if other operators are measured.

\subsubsection{Measuring the variance of an angular momentum component}

A realistic method to model the imperfections
is by introducing local decoherence channels for each qubit as
 \begin{eqnarray}
\epsilon_q^{({\rm wn})}(\varrho)=(1-q)\varrho+q\frac{\openone}{2},
\end{eqnarray}
where $0\le q\le1.$
Each spin is mixed with a certain amount of white noise locally.
The above single-qubit decoherence is given by the Kraus operators as
$\epsilon_q^{({\rm wn})}(\varrho) = (1-q) \varrho + {q}\sum_i   K_i \varrho K_i^\dagger $  where ${\bf K}=\{\openone,\sigma_x,\sigma_y,\sigma_z \}/2$. 

We assume that all these decoherence channels act in parallel for all spins. Thus, 
we obtain the multi-spin decoherence
 \begin{eqnarray}
 \mathcal{E}_q^{({\rm wn})}(\varrho)=\epsilon_q^{({\rm wn},1)}(\varrho)\circ 
\epsilon_q^{({\rm wn},2)}(\varrho)\circ ... \circ \epsilon_q^{({\rm wn},N)}(\varrho). \label{EEE}\end{eqnarray}
In Eq.~(\ref{EEE}), $\circ$ indicates function composition $(f\circ g)(x) = f(g(x))$ and the number in the superscript indicates to which qubit the function is applied.
We consider an atomic ensemble with a continuous distribution.
Next we outline briefly the derivation of the noisy dynamics for this case.

The moments $\ex{\Jc^2}$ and $\ex{\Jc^4}$ given in Eqs.~(\ref{Jx2b}) and 
 (\ref{Jx4}),
 respectively, are affected through the two-body and four-body correlations via the formulas
\bea
\sum_{n_1 \neq n_2}\ex{\JJc_{\Theta}^{(n_1)}\JJc_{\Theta}^{(n_2)}}_{\varrho_{\rm s,wn}}=(1-q)^2
\sum_{n_1 \neq n_2}\ex{\JJc_{\Theta}^{(n_1)}\JJc_{\Theta}^{(n_2)}}_{\varrho_{\rm s}}\nonumber\\
\eea
and
\bea
&&\sum_{\neq (n_1,n_2,n_3,n_4)}\ex{\JJc_{\Theta}^{(n_1)}\JJc_{\Theta}^{(n_2)}\JJc_{\Theta}^{(n_3)}\JJc_{\Theta}^{(n_4)}}_{\varrho_{\rm s,wn}}\nonumber\\
&&\;\;\;\;\;\;=(1-q)^4\sum_{\neq (n_1,n_2,n_3,n_4)}\ex{\JJc_{\Theta}^{(n_1)}\JJc_{\Theta}^{(n_2)}\JJc_{\Theta}^{(n_3)}\JJc_{\Theta}^{(n_4)}}_{\varrho_{\rm s}}.\nonumber\\
\eea
Hence, for the atomic cloud we obtain (cf. Observation \actualize{5 and 6}{5 and 6})
\begin{equation}
\ex{\Jc^2}^{f_N^\indfprod}_{{\rm s,wn}}(q) =\frac{N\hbar^2}{4}\Big[1-(1-q)^2(\tilde C^2+\tilde S^2)\Big]\label{angularmomentum2bp}
\end{equation}
and
\bea
\ex{\Jc^4}^{f_N^\indfprod}_{{\rm s,wn}}(q) &=& 
\frac{N\hbar^4}{16}\bigg\{
3N-2
-(1-q)^2(6N-8)(\tilde{C}^2+\tilde{S}^2)\nonumber \\ 
&& \qquad\quad + (1-q)^4 3(N-2)(\tilde{C}^2+\tilde{S}^2)^2 \bigg\}. \label{MomAng_dp}
\eea
Note that for $q=1,$ Eqs.~(\ref{angularmomentum2bp}) and (\ref{MomAng_dp}) reduce to the values corresponding to a global white noise. 
Equation (\ref{angularmomentum2bp}) can be rewritten as
\begin{eqnarray}
\ex{\Jc^2}_{\rm s,\indwn}(q)&=&
(1-q)^2 \ex{\Jc^2}_{\rm s}+
[1-(1-q)^2]\ex{\Jc^2}_\indwn.\nonumber\\
\end{eqnarray}
Let us consider a Gaussian density profile described in Sec.~\ref{SecGaussianDensityProf} for which we have $\tilde{C}^2+\tilde{S}^2$ given in Eq.~(\ref{CSGaussian}).
Substituting Eq.~(\ref{CSGaussian}) into Eqs.~(\ref{angularmomentum2bp}) and (\ref{MomAng_dp}),
we can compute the influence of the noise for the second and fourth moments of the angular momentum components.
The precision of the reconstruction of $\Theta$ based on measuring $\Jc^2$ can be obtained as
\begin{equation}
(\Delta \Theta)^{2}_{\rm s,\indwn} =  \frac{\ex{J_x^4}_{\rm s,\indwn}(\Theta)-\ex{J_x^2}_{\rm s,\indwn}
(\Theta)^2}{(1-q)^4|\partial_{\Theta}\ex{J_x^2}_{\rm s}|^2} , \label{DeltaTheta_n}
\end{equation}

Let us see now the behavior of the precision given by Eq.~(\ref{DeltaTheta_n}) under noise in the limiting 
cases, {for $q>0$. }

(i) At  $\Theta=0,$ the {denominator} of the right-hand side of Eq.~(\ref{DeltaTheta_n}) is zero.
{This can be seen from Eq.~(\ref{eq:Jx2_equidist_chain}), from which
it follows that $\partial_\Theta\ex{J_x^2}_{\rm s}$
is a sum of sine expressions, which vanish at $\Theta=0$.
In contrast,} 
the {numerator} is a positive number. Hence, even for very small amount of noise we have
\begin{equation}
(\Delta \Theta)^{-2}_{\rm s,\indwn} (\Theta=0) = 0.
\end{equation}
{Note that this is due to the fact that we chose a noise state which is invariant
under the $\Theta$ dependent transformation $U_{\rm G}$ from Eq.~(\ref{eq:Ugradient}).}
The precision remains close to zero until the noise of the singlet becomes comparable to
the added noise.

(ii) The other limit is the case of the large $\Theta.$ 
Let us define $\Theta_\indwn$ such that for $\Theta > \Theta_\indwn$ the singlet evolved into a state
that, based on the second and fourth moments of the angular momentum coordinates, is like the completely mixed state.
That is, for $\Theta > \Theta_\indwn$ we have
\bea
\ex{\Jc^2}_{\rm s} (\Theta) &\approx& \ex{\Jc^2}_\indwn,\nonumber\\
\ex{\Jc^4}_{\rm s} (\Theta) &\approx& \ex{\Jc^4}_\indwn.
\eea
The reason for this is that for large enough $\Theta$ the quantity  $\tilde{C}^2+\tilde{S}^2$ as given in Eq.~(\ref{CSGaussian}) is close to zero and in Eqs.~(\ref{angularmomentum2bp}) and (\ref{MomAng_dp}) only the constant terms corresponding to the moments of the  white noise remain.
Hence, based on Eq.~(\ref{DeltaTheta_n}) for  $\Theta>\Theta_\indwn,$ we have 
\bea
(\Delta \Theta)^{-2}_{\rm s,\indwn} (\Theta)\approx(1-q)^4 (\Delta \Theta)^{-2}_{\rm s} (\Theta).  \label{DeltaTheta_n2}
\eea

\begin{figure}%[h!]
\centering
           \includegraphics[width=0.50\textwidth]{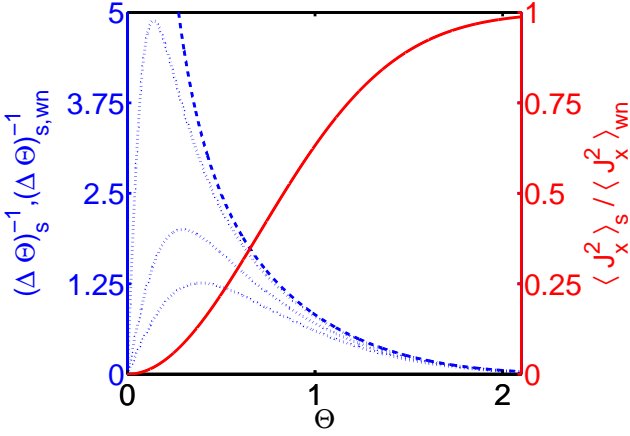}
           \vskip -0.9cm
\caption{(Color online) The dynamics of $\ex{\Jc^2}_{\rm s}/\ex{\Jc^2}_\indwn$ (solid line) and 
the precision $(\Delta \Theta)_{\rm s}^{-1}$ for the noiseless case (dashed line) and $(\Delta \Theta)_{\rm s,wn}^{-1}$ for 
a noise with $q=0.01,0.05$ and 0.1 (from top to bottom, dotted line) as the function of $\Theta$ for $N=10^5$ particles. $\ex{\Jc^2}_\indwn$ is defined in Eq.~(\ref{Jc_cm}).
The density profile is Gaussian and $\sigma=L.$
} 
\label{fig_Jx2_Fisher_Gauss_noise}
 \end{figure}

In Fig.~\ref{fig_Jx2_Fisher_Gauss_noise}, we calculated the precision $(\Delta\Theta)_{\rm s,wn}(\Theta)$ as a function of the noise for 
$q=0.01,0.05$ and $0.1$ for a Gaussian ensemble.
For these values, for large $\Theta$ we have $(\Delta \Theta)^{-2}_{\rm s} (\Theta)/ (\Delta \Theta)^{-2}_{\rm s,\indwn}(\Theta)=1.04,1.23,$ and $1.52,$ respectively.

\subsubsection{Measuring operators different from the angular momentum components}

We will show a simple example that even in the noisy case a higher
accuracy of the gradient estimation can be achieved if quantities
other than $\langle J_{x}^{2}\rangle$ are measured. Let us consider
a noisy singlet of the type
\begin{equation}
\varrho_{{\rm ns}}=p_{{\rm n}}\frac{\openone}{2^{N}}+(1-p_{{\rm n}})\varrho_{{\rm s}}.
\end{equation}
Let us now look at the projector to the $J_{x}=0$ subspace. For the
completely mixed state, the expectation value of the projector is
\begin{equation}
\langle P_{J_{x}=0}\rangle_{{\rm n}}=2^{-N}\binom{N}{\frac{N}{2}}\approx\sqrt{\frac{2}{\text{\ensuremath{\pi}}}}\frac{1}{\sqrt{N}},
\end{equation}
while for the singlet we have $\langle P_{J_{x}=0}\rangle_{{\rm s}}(\Theta=0)=1.$ Hence, for the noisy state we obtain
\begin{equation}\label{proj}
\langle P_{J_{x}=0}\rangle_{{\rm ns}}\approx p_{{\rm n}}
\sqrt{\frac{2}{\text{\ensuremath{\pi}}}}\frac{1}{\sqrt{N}}+(1-p_{{\rm n}})\langle P_{J_{x}=0}\rangle_{{\rm s}}.
\end{equation}
During the dynamics, based on Eq.~(\ref{proj}),  the expectation value of the projector $\langle P_{J_{x}=0}\rangle_{{\rm s}}$ decreases from $1$ to a value close to zero, while the noise in the expectation value of the projector  $\langle P_{J_{x}=0}\rangle_{{\rm n}}$ is proportional to $N^{-\frac{1}{2}},$ i.e., it is
$O(N^{-\frac{1}{2}}).$ 
In contrast, if the second moment $\langle J_x^2\rangle$ is measured  then 
\begin{equation}\label{secondm}
\langle J_x^2 \rangle_{{\rm ns}} = p_{{\rm n}} \frac{N\hbar^2}{4}+(1-p_{{\rm n}})\langle J_x^2 \rangle_{{\rm s}}.
\end{equation}
Based on Eq.~(\ref{secondm}), 
$\langle J_x^2 \rangle_{{\rm s}}$ changes from $0$ to a value close to $O(N),$ while the noise in the expectation value of the second moment $\langle J_x^2 \rangle_{{\rm n}}$ is of the same order, $O(N).$ 
It can be seen that  the effect of the noise is much smaller in the expectation  value of the projector (\ref{proj}) than in the second moment (\ref{secondm}). An analogous calculation can be carried out for local noise
channels for operators of the type $P_{J_x=\text{const.}}.$ 

%%%%%%%%%%%%%%%%%%%%%%%%%%%%%%%
\subsection{Spin-$j$ particles}
%%%%%%%%%%%%%%%%%%%%%%%%%%%%%%%
\label{Sec_Largej}

So far we have discussed the case of singlet states of $j=\frac{1}{2}$ particles.
In this section we will study the singlet of spin-$j$ particles. We will find that the dynamics of $\ex{\Jc^2}$ for $N$ spin-$j$ particles is the same as the dynamics of $\ex{\Jc^2}$ for $N$ spin-$\frac{1}{2}$ particles, when in both cases we normalize the variance with that of the white noise. 
Moreover, 
{we find that by using a certain Gaussian assumption in order 
to estimate $\ex{J_x^4}_{\rm s}$,} 
we obtain the same dynamics even for the precision.

When considering the spin-$j$ case, one could think about 
using the ideas of Observation 1 to obtain the multiparticle singlet of spin-$j$ particles 
as an equal mixture of all the permutations
of tensor products of two-particle singlets. However, the state obtained this way does 
not equal Eq.~(\ref{singlet_J0}).
This is due to the fact that for $j>\frac{1}{2},$ there are several permutationally invariant $SU(2)$ 
singlets \cite{SUdsinglet}.
Hence, for $j>\frac{1}{2},$ another method is needed.

For the dynamics of the variance of the collective angular momentum components,
we need again the variance for the completely mixed state (white noise), which is obtained for $N$ spin-$j$ particles as
\bea
\ex{\Jc^2}_{\indwn,j}=\frac{Nj(j+1)}{3}\hbar^2,
\eea
Note that $\ex{\Jc^2}_{\indwn,\frac{1}{2}}=\ex{\Jc^2}_{\indwn}$ where $\ex{\Jc^2}_{\indwn}$ is defined in 
Eq.~(\ref{Jc_cm}). As we will see, the dynamics of the variance of the singlet will be similar to the dynamics for the $j=\frac{1}{2}$ case, apart from a factor of  
\bea
\kappa_j=\frac{\ex{\Jc^2}_{\indwn,j}}{\ex{\Jc^2}_{\indwn}}=\frac{4}{3}j(j+1)
\eea
 and hence for large $\Theta$ we will have  $\ex{\Jc^2}_{{\rm s},j}(\Theta)\approx\ex{\Jc^2}_{\indwn,j}.$

Let us denote by $j_l$ the angular momentum components of a single spin-$j$ particle.
Since for each spin the dynamics is invariant under a coordinate transformation, from 
$(j_x^{(n)})^2+(j_y^{(n)})^2+(j_z^{(n)})^2=j(j+1)\hbar^2$ we obtain
 \begin{equation}\label{eq:Jx2j1}
 \ex{\sum_n (j_l^{(n)})^2}(\Theta) =\ex{\Jc^2}_{\indwn,j},
 \end{equation}
for $l=x,y,z$ and for all $\Theta.$ 
Moreover, from the requirement that $\ex{J_l^2}_{{\rm s},j}=0,$ we obtain
\bea
\ex{\Jc^2}=\ex{\sum_n (j_z^{(n)})^2}_{{\rm s},j}+N(N-1)\ex{j_l^{(1)}j_l^{(2)}}_{{\rm s},j}=0.\nonumber\\
\eea
{Note that $\ex{j_l^{(1)}j_l^{(2)}}_{{\rm s},j}=\ex{j_l^{(n)}j_l^{(n')}}_{{\rm s},j}$
for $n\neq n'$ due to the permutational invariance of the state.}
Hence we arrive at
\bea\label{eq:Jx2j2}
\ex{j_l^{(n)}j_l^{(n')}}_{{\rm s},j} = -\frac{\ex{\Jc^2}_{\indwn,j}}{N(N-1)},
\eea
and we can rewrite Eq.~(\ref{deltakl}) for particles with $j>\frac{1}{2}$ as 
\bea
\ex{j_k\otimes j_l}_{\varrho_{12}^{{\rm red},j}}=-\frac{\ex{\Jc^2}_{\indwn,j}}{N(N-1)}\delta_{kl}, \label{deltakl2}
\eea
where $\varrho_{12}^{{\rm red},j}$ is now the reduced state of the spin-$j$ singlet
{\cite{Note_SU2invariance}}.

{The time-evolved single particle operators $j_x(\Theta)\equiv X(\Theta)$ 
have the same form as in the $j=\frac{1}{2}$ case, given in Eq.~(\ref{eq:Xop}).
Therefore, we can perform the calculation for any $j$ in analogy to the 
derivation for the spin-$\frac{1}{2}$ case of Sec. III.A. Starting from 
}
\bea
\ex{\Jc^2(\Theta)}_{{\rm s},j}=\ex{\Jc^2}_{\indwn,j}+\sum_{n_1 \neq n_2} \ex{\JJc_{\Theta}^{(n_1)}\JJc_{\Theta}^{(n_2)}}_{{\rm s},j},
\label{Jx2b1}
\eea
{we arrive at}
\be
[ \ex{\Jc^2}(\Theta)]_{{\rm s},j}=\kappa_j[ \ex{\Jc^2}(\Theta)]_{{\rm s}},
 \label{FormulaLargaj1b}
\ee
{Therefore, Observation 2 generalizes to spin-$j$ particles as follows.}

{\bf Observation 8.}  
{\it 
The dynamics of the variance of $\Jc$ for a chain of spin-$j$ particles with
the $z$-coordinates $\vec z_N^\indchainV$ is
\bea
&& \ex{\Jc^2}_{{\rm s},j}^{\vec z_N^\indchainV}(\Theta)= \kappa_j \frac{\hbar^2 N }{4} \bigg\{1+\frac{1}{N(N-1)}[N-\CTheta^2-\STheta^2]\bigg\}\nonumber\\
 \label{FormulaLargaj1}
\eea
where $C$ and $S$ are defined in Eq.~(\ref{CS}).
}

For an ensemble of particles with a density profile 
$\lambda(z),$ we can in analogy generalize Observation 5 to any $j$
as follows. 

{
{\bf Observation 9.}
{\it
For an ensemble of spin-$j$ particles with a 
product distribution function $f_N^\indfprod$ from Eq.~(\ref{uncorr}),
we obtain the following dynamics for the expectation value of 
the second moment of $\Jc$:
\begin{equation}
\ex{\Jc^2}^{f_N^\indfprod}_{\rm s}(\Theta)=\kappa_j \frac{N\hbar^2}{4}\Big[1-\tilde C^2-\tilde S^2\Big],
\label{angularmomentum21}
\end{equation}
where $\tilde{C}$ and $\tilde{S}$ are defined in  Eqs.~(\ref{CT}) and (\ref{ST}). Here the averaging is over 
the density distribution $f_1(z_1).$
}
}

For the Gaussian density profile (\ref{GaussianDensity}), we arrive at
\begin{equation}
 \ex{\Jc^2}(\Theta)=\kappa_j \frac{N}{4}\hbar^2\bigg(1-e^{-\frac{\sigma^2}{L^2} \Theta^2}\bigg).
\end{equation}

The calculation of $\ex{\Jc^4}(\Theta)$ for the singlet of spin-$j$ particles seems to be much more complicated than for spin-$\frac{1}{2}$ particles. 
%----------------------[GAUSSIANITY]-----------------------------%
It is possible to avoid calculating the fourth-order moment by using the assumption that 
when $\Jc$ is measured, the probability of the measurement outcomes follow a Gaussian curve,
with the zero outcome being the most probable. For such Gaussian probability distributions, the 
higher-order moments can be obtained from second-order ones \cite{Gaussian}. In particular, for our case,
\begin{eqnarray}
\ex{\Jc^4}\approx 3\ex{\Jc^2}^2. \label{exJ4_Gaussian}
\end{eqnarray}
Equation~(\ref{exJ4_Gaussian}) leads to $\va{\Jc^2}\approx 2\ex{\Jc^2}^2,$
keeping in mind that $\ex{\Jc}=0.$ Such a Gaussianity assumption
is expected to work  for later times, as at $\Theta=0$ the variance of $\Jc$
is zero, and in the beginning only a few of the eigenstates of $\Jc$ are populated.
Later, however, many eigenstates of $\Jc$ are populated and a continuous approximation
of the discrete spectrum is appropriate. Note that the Gaussianity of the probability distribution
is a notion completely independent from the Gaussianity of the density profile of the cold gas. 

One can substitute the Gaussian assumption (\ref{exJ4_Gaussian}) into the formula 
for $(\Delta \Theta)_{\rm s}^2$ given in Eq.~(\ref{DeltaTheta}) {which 
can be used for any $j$.}
This clearly simplifies the calculations.
For the accuracy of gradient metrology 
for $N$ spin-$j $ particles we obtain the same result as for $N$ spin-$\frac{1}{2}$ particles
\begin{equation}
(\Delta\Theta)_{{\rm s},j}(\Theta)=(\Delta\Theta)_{{\rm s},\frac{1}{2}}(\Theta).
\end{equation}

\begin{figure}%[h!]
\centering
           \includegraphics[width=0.45\textwidth]{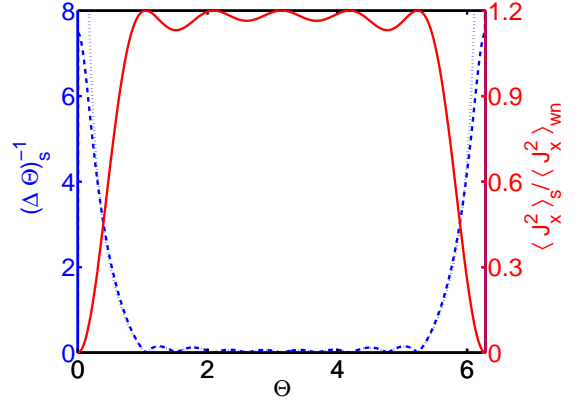}
           \vskip -0.5cm
\caption{(Color online) The dynamics of $\ex{\Jc^2}_{\rm s}/\ex{\Jc^2}_\indwn$ (solid line) and $(\Delta \Theta)_{\rm s}^{-1}$ (dashed line) as the function of $\Theta$ for $N=6$ spin-$1$ particles in an equidistant chain.
We also present $(\Delta \Theta)_{\rm s}^{-1}$ calculated based on the Gaussianity assumption 
given by Eq.~(\ref{exJ4_Gaussian}) discussed in Sec.~\ref{Sec_Largej} (dotted line).
}
\label{fig_j1_Gauss}
 \end{figure}

In Fig.~\ref{fig_Jx2_Fisher_Gauss}, besides the exact result,
we also plot the dynamics based on the approximation using the Gaussian 
assumption on the correlations given by Eq.~(\ref{exJ4_Gaussian}) {for the 
$j=\frac{1}{2}$ case as an illustration}.
The bound obtained this way diverges at $t=0,$
however, it is very close to the true value for $\Theta >0.025.$
We also carried out a calculation for $j>\frac{1}{2}.$
In Fig.~\ref{fig_j1_Gauss},  one can see the comparison
for the case of a chain of six spin-$1$ particles. The precision based on the Gaussian
assumption diverges at $t=0;$ however, later it fits 
the exact dynamics very well.

%----------------------[GAUSSIANITY]-----------------------------%

%%%%%%%%%%%%%%%%%%%%%%%%%%%%%%%
\section {Discussion and Conclusion}
\label{sec:disc}
%%%%%%%%%%%%%%%%%%%%%%%%%%%%%%%

\comment{We might need to improve this further!}

We presented calculations for a many-particle singlet in a magnetic field gradient
for an ensemble of distinguishable, well-localized spin-$\frac{1}{2}$ 
particles. 
We have shown that multiparticle singlets created in cold atomic ensembles can be used for 
differential magnetometry. The magnetic field gradient can be estimated by measuring the 
variance of one of the components of the collective angular momentum.
We calculated the dependence of this variance on the field gradient and 
the measurement time. We also calculated the precision of the estimation
of the field gradient. {We have also considered admixtures of white noise and discussed 
the extension of the results for general spin-$j$ particles.}
Our work opens up the possibilities for experiments
with unpolarized ensembles.

In the future, it would be interesting to find bounds for the precision of the field gradient 
measurements using the theory of the quantum Fisher information \cite{Fisher}.
It is an interesting question as to whether the accuracy of parameter estimation in our calculations
saturates the accuracy bound determined by the quantum Fisher information.
It would also be important to consider the case of particles that are not well localized,
in which case the spatial degree of freedom cannot be easily separated from the internal degrees
of freedom for the quantum dynamics we considered \cite{Fisher}.
For room-temperature experiments with atomic ensembles, 
the atoms must be well localized; however, singlets can also be realized with Bose-Einstein condensates,
for which all particles are delocalized. For such systems, the noise is very different from the case
of distinguishable particles. Finally, the effect of a field gradient could be used to 
examine symmetric Dicke states with $\langle J_z \rangle=0$  rather than singlets \cite{LS12,Hamley2012}. 
Our calculations could be generalized to that case.

%%%%%%%%%%%%%%%%%%%%%%%%%%%%%%%
%\section {Conclusions}
%%%%%%%%%%%%%%%%%%%%%%%%%%%%%%%

\begin{acknowledgments}
We thank J.~Calsamiglia, G.~Colangelo, O.~G\"uhne, M.~Modugno, L.~Santos, R.J.~Sewell, and  Z.~Zimbor\'as for discussions. 
We thank the European Union (ERC Starting
Grants GEDENTQOPT and AQUMET, CHIST-ERA
QUASAR), the Spanish MINECO (Projects No.
FIS2009-12773-C02-02, FIS2012-36673-C03-03 and FIS2011-23520) the Basque
Government (Projects No. IT4720-10 and IT559-10), and the support
of the National Research Fund of Hungary OTKA (Contract
No. K83858). I.U.L. acknowledges the support of
a Ph.D. grant of the Basque Government.
\end{acknowledgments}

%%%%%%%%%%%%%%%%%%%%%%%%%%%%%%%%%%%%%%%%%%%%%%%%%%%%%%%%%%%%%%%%%%%%%%

% \end{document}

%%%%%%%%%%%%%%%%%%%%%%%%%%%%%%%%%%%%%%%%%%%%%%%%%%%%%%%%%%%%%%%%%%

% taken from singlet_EPAPS8e2_Geza.tex

\onecolumngrid

\newpage

\renewcommand{\thefigure}{S\arabic{figure}}
\renewcommand{\thetable}{S\arabic{table}}
\renewcommand{\theequation}{S\arabic{equation}}
\setcounter{figure}{0}
\setcounter{table}{0}
\setcounter{equation}{0}
\setcounter{section}{0}
\setcounter{page}{1}
\newcommand{\loo}{{\lambda}}

\newcommand{\proofend}{\hfill\fbox\\\medskip }

\section*{\large Macroscopic singlet states for gradient magnetometry (Supplemental
Material)}
\begin{center}
I\~nigo Urizar-Lanz,$^1$ Philipp Hyllus,$^1$ I\~nigo Luis Egusquiza,$^1$ Morgan W. Mitchell,$^{2,3}$ and G\'eza T\'oth$^{1,4,5}$
\vskip0.5cm
{\it \small
$^1$Department of Theoretical Physics, University of the Basque Country UPV/EHU, P.O. Box 644, E-48080
Bilbao, Spain

$^2$ICFO-Institut de Ciencies Fotoniques, Mediterranean Technology Park, E-08860 Castelldefels (Barcelona), Spain

$^3$ICREA-Instituci\'{o} Catalana de Recerca i Estudis Avan\c{c}ats, E-08015 Barcelona, Spain

$^4$IKERBASQUE, Basque Foundation for Science, E-48011 Bilbao, Spain

$^5$Wigner Research Centre for Physics, Hungarian Academy of 
Sciences, P.O. Box 49, H-1525 Budapest,  Hungary
}
\vskip0.5cm
\end{center}

\hskip2cm In the supplement, we present additional calculations for obtaining the accuracy of 

\hskip2cm the gradient measurement.

%-------------------------------------------------------------------------------%
\section{Additional calculations for obtaining $\langle J_{x}^{4}\rangle_{\rm s}(\Theta)$
from Observation 3}

The general expression of $\langle J_x^4\rangle_{\rm s}(\Theta)$ for $N$
particles being at the positions $z_k^\indchain$, collected in a vector $\vec z_N^\indchainV$,
has been obtained in \actualize{Eq.~(60)}{Eq.~(\ref{jx4X})}. 
Here and in the following we will label with ${\vec z}_N^\indchainV$ a vector
with $N$ elements $z_1^\indchain,...,z_N^\indchain$. 
In order to being able to compute this for large $N$, we need to simplify
\begin{equation}
I_4:=I_{{\rm cccc}}+I_{{\rm ssss}}+2I_{{\rm ccss}},\label{eq:S}
\end{equation}
 where 
\begin{align}
I_{{\rm cccc}} & :=\sum_{\ne(k,l,m,n)}c_{k}c_{l}c_{m}c_{n},\nonumber \\
I_{{\rm ssss}} & :=\sum_{\ne(k,l,m,n)}s_{k}s_{l}s_{m}s_{n},\nonumber \\
I_{{\rm ccss}} & :=\sum_{\ne(k,l,m,n)}c_{k}c_{l}s_{m}s_{n},
\end{align}
where $c_k=\cos(\frac{z_k^\indchain}{L}\Theta)$ and $s_k=\sin(\frac{z_k^\indchain}{L}\Theta)$.
We will show two ways of doing this. Firstly, as stated in \actualize{Eq.~(61)}{Eq.~(\ref{III444})},
one can rewrite it as 
\begin{eqnarray}
I_4 & = & X_{1,0}^{4}+X_{0,1}^{4}+2X_{1,0}^{2}X_{0,1}^{2}\nonumber \\
 &  & -(6X_{2,0}X_{1,0}^{2}+6X_{0,2}X_{0,1}^{2}+2X_{2,0}X_{0,1}^{2}+2X_{0,2}X_{1,0}^{2}+8X_{1,1}X_{1,0}X_{0,1})\nonumber\\
 &  & +8X_{3,0}X_{1,0}+3X_{2,0}^{2}+8X_{0,3}X_{0,1}+3X_{0,2}^{2}+8X_{2,1}X_{0,1}+2X_{2,0}X_{0,2}+8X_{1,2}X_{1,0}+4X_{1,1}^{2}\nonumber \\
 &  & -(6X_{4,0}+6X_{0,4}+12X_{2,2}),\label{eq:statement1}
\end{eqnarray}
 where 
\begin{equation}
X{}_{m,n}:=\sum_{k=1}^{N}c_{k}^{m}s_{k}^{n}.
\end{equation}
The proof is presented in Section~\ref{sec:Q1} below. 
Secondly, as stated in  \actualize{Eq.~(63)}{Eq.~(\ref{III444B})}, one may also rewrite it more compactly as 
\begin{equation}
I_4= N \left\{2(N-3)-4 N(N-2)\left|\hat{f}_1(\alpha)\right|^2+N^3\left|\hat{f}_1(\alpha)\right|^4 + 
N\left|\hat{f}_1(2\alpha)\right|^2-2N^2\textrm{Re}\big[\hat{f}_1^2(\alpha)\hat{f}_1(2 \alpha)^*\big]\right\}
,\label{eq:statement2}
\end{equation}
where 
\begin{equation}
	\hat f_1(\alpha)=\frac{1}{N}\sum_k e^{i\alpha z_k^\indchain}\quad \mathrm{and}\quad
	\alpha=\frac{\Theta}{L}.
	%\langle e^{i\alpha z_1}\rangle=\int {\rm d}z_1 f_1(z_1)e^{i\alpha z_1}.
\end{equation}
The proof is presented in Section~\ref{sec:Q2} below.

%- - - - - - - - - - - - - - - - - - - - - - - - - - - - - - - - - - - - - - - -%
\subsection{Proof of Eq.~(\ref{eq:statement1})}
\label{sec:Q1}

Let us concentrate on the last term in Eq.~\eqref{eq:S}.
We can write using a shorthand notation
\begin{align}
I_{{\rm ccss}} & =\left(\sum_{k,l,m,n}\right.\nonumber \\
 & -\sum_{\ne(k=l,m,n)}%\nonumber \\& 
 -\sum_{\ne(k,l,m=n)}%\nonumber \\& 
 -\sum_{\ne(k=m,l,n)}-\sum_{\ne(k=n,m,n)}-\sum_{\ne(k,l=m,n)}-\sum_{\ne(k,l=n,m)}\nonumber \\
 & -\sum_{\ne(k=l,m=n)}%\nonumber \\& 
 -\sum_{\ne(k=m,l=n)}-\sum_{\ne(k=n,m=n)}\nonumber \\
 & -\sum_{\ne(k=l=m,n)}-\sum_{\ne(k=l=n,m)}%\nonumber \\& 
 -\sum_{\ne(k=m=n,l)}-\sum_{\ne(l=m=n,k)}\nonumber \\
 & \left.-\sum_{k=l=m=n}\right)c_{k}c_{l}s_{m}s_{n}.\label{eq:ccss}
\end{align}
 Here $\ne(k=m,l=n)$ means that the summation is such that $k=m,l=n$
and $k\ne l.$ Eq.~\eqref{eq:ccss} can be rewritten after simple
considerations as
\begin{eqnarray}
I_{{\rm ccss}} & = & \sum_{k,l,m,n}c_{k}c_{l}s_{m}s_{n}-\sum_{\ne(k,l,m)}(c_{k}^{2}s_{l}s_{m}+c_{k}c_{l}s_{m}^{2}+4c_{k}s_{k}c_{l}s_{m})\nonumber \\
 &  & -\sum_{\ne(k,l)}(c_{k}^{2}s_{l}^{2}+2c_{k}s_{k}c_{l}s_{l}+2c_{k}^{2}s_{k}c_{l}+2c_{k}c_{l}s_{l}^{2})-\sum_{k}c_{k}^{2}s_{k}^{2}.\label{eq:ccss2}
\end{eqnarray}
Each term in Eq.~\eqref{eq:ccss2} corresponds to a line in Eq.~\eqref{eq:ccss}.
Then, we can rewrite the terms in Eq.~\eqref{eq:ccss2} still containing
the conditions ``not equal'' with terms without such conditions
as follows 
\begin{align}
\sum_{\ne(k,l,m)}c_{k}^{2}s_{l}s_{m}= & \left(\sum_{k,l,m}-\sum_{\ne(k=l,m)}-\sum_{\ne(k,l=m)}-\sum_{\ne(k=m,l)}-\sum_{k=m=n}\right)c_{k}^{2}s_{l}s_{m}\nonumber \\
= & \sum_{k,l,m}c_{k}^{2}s_{l}s_{m}\nonumber \\
 & -2(\sum_{k,l}c_{k}^{2}s_{k}s_{l}-\sum_{k}c_{k}^{2}s_{k}^{2})-(\sum_{k,l}c_{k}^{2}s_{l}^{2}-\sum_{k}c_{k}^{2}s_{k}^{2})\nonumber \\
 & -\sum_{k}c_{k}^{2}s_{k}^{2},\nonumber \\
= & \sum_{k,l,m}c_{k}^{2}s_{l}s_{m}-\sum_{k,l}(2c_{k}^{2}s_{k}s_{l}+c_{k}^{2}s_{l}^{2})+2\sum_{k}c_{k}^{2}s_{k}^{2},\label{eq:c2ss}
\end{align}
 where we used that we have 
\begin{equation}
\sum_{\ne(k,l)}a_{k}b_{l}=\sum_{k,l}a_{k}b_{l}-\sum_{k}a_{k}b_{k}\label{eq:2ind}
\end{equation}
 for any real numbers $a_{k}$ and $b_{k}.$ Analogously, one finds
that 
\begin{equation}
\sum_{\ne(k,l,m)}c_{k}c_{l}s_{m}^{2}=\sum_{k,l,m}c_{k}c_{l}s_{m}^{2}-\sum_{k,l}(2c_{k}c_{l}s_{l}^{2}+c_{k}^{2}s_{l}^{2})+2\sum_{k}c_{k}^{2}s_{k}^{2},\label{eq:ccs2}
\end{equation}
 and 
\begin{equation}
\sum_{\ne(k,l,m)}c_{k}s_{k}c_{l}s_{m}=\sum_{k,l,m}c_{k}s_{k}c_{l}s_{m}-\sum_{k,l}(c_{k}^{2}s_{k}s_{l}+c_{k}s_{k}c_{l}s_{l}+c_{k}c_{l}s_{l}^{2})+2\sum_{k}c_{k}^{2}s_{k}^{2}.\label{eq:cscs}
\end{equation}
Substituting Eqs.~\eqref{eq:c2ss}, \eqref{eq:ccs2}, and \eqref{eq:cscs}
into Eq.~\eqref{eq:ccss2}, and using again Eq.~\eqref{eq:2ind}
for the remaining terms of two non-equal indices, we arrive at 
\begin{align}
I_{{\rm ccss}}= & \sum_{k,l,m,n}c_{k}c_{l}s_{m}s_{n}\nonumber \\
 & -\sum_{k,l,m}c_{k}^{2}s_{l}s_{m}+\sum_{k,l}(2c_{k}^{2}s_{k}s_{l}+c_{k}^{2}s_{l}^{2})-2\sum_{k}c_{k}^{2}s_{k}^{2}\nonumber \\
 & -\sum_{k,l,m}c_{k}c_{l}s_{m}^{2}+\sum_{k,l}(2c_{k}c_{l}s_{l}^{2}+c_{k}^{2}s_{l}^{2})-2\sum_{k}c_{k}^{2}s_{k}^{2}\nonumber \\
 & -4\sum_{k,l,m}c_{k}s_{k}c_{l}s_{m}+4\sum_{k,l}(c_{k}^{2}s_{k}s_{l}+c_{k}s_{k}c_{l}s_{l}+c_{k}c_{l}s_{l}^{2})-8\sum_{k}c_{k}^{2}s_{k}^{2}\nonumber \\
 & -\sum_{k,l}(c_{k}^{2}s_{l}^{2}+2c_{k}s_{k}c_{l}s_{l}+2c_{k}^{2}s_{k}s_{l}+2c_{k}c_{l}s_{l}^{2})+7\sum_{k}c_{k}^{2}s_{k}^{2}\nonumber \\
 & -\sum c_{k}^{2}s_{k}^{2}.\label{eq:ccss3}
\end{align}
 In Eq.~\eqref{eq:ccss3}, the first four lines correspond to the
first line in Eq.~\eqref{eq:ccss2}, and the remaining two lines
to the second line. This can be simplified by combining terms that
appear more than once as 
\begin{align}
I_{{\rm ccss}}= & \sum_{k,l,m,n}c_{k}c_{l}s_{m}s_{n}\nonumber \\
 & -\sum_{k,l,m}\left(c_{k}^{2}s_{l}s_{m}+c_{k}c_{l}s_{m}^{2}+4c_{k}s_{k}c_{l}s_{m}\right)\nonumber \\
 & +\sum_{k,l}\left(c_{k}^{2}s_{k}s_{l}(2+4-2)+c_{k}^{2}s_{l}^{2}(1+1-1)+c_{k}c_{l}s_{l}^{2}(2+4-2)+c_{k}s_{k}c_{l}s_{l}(4-2)\right)\nonumber \\
 & +\sum_{k}c_{k}^{2}s_{k}^{2}(-2-2-8+7-1),\label{eq:ccss3-1}
\end{align}
 which finally yields 
\begin{eqnarray}
I_{{\rm ccss}} & = & \sum_{k,l,m,n}c_{k}c_{l}s_{m}s_{n}-\sum_{k,l,m}\left(c_{k}^{2}s_{l}s_{m}+c_{k}c_{l}s_{m}^{2}+4c_{k}s_{k}c_{l}s_{m}\right)\nonumber \\
 &  & +\sum_{k,l}\left(4c_{k}^{2}s_{k}s_{l}+c_{k}^{2}s_{l}^{2}+4c_{k}c_{l}s_{l}^{2}+2c_{k}s_{k}c_{l}s_{l}\right)-6\sum_{k}c_{k}^{2}s_{k}^{2}.\label{eq:ccss3-1-1}
\end{eqnarray}
 The formula for $I_{{\rm cccc}}$ can be obtained from the formula
for $I_{{\rm ccss}}$ {[}Eq.~\eqref{eq:ccss3-1-1}{]}, by replacing
$s_{k}$ by $c_{k}$ and combining terms that appear more than once
as 
\begin{align}
I_{{\rm cccc}} & =\sum_{k,l,m,n}c_{k}c_{l}c_{m}c_{n}-\sum_{k,l,m}c_{k}^{2}c_{l}c_{m}(1+1+4)+\sum_{k,l}c_{k}^{3}c_{l}(4+4)+\sum_{k,l}c_{k}^{2}c_{l}^{2}(1+2)-6\sum_{k}c_{k}^{4}\nonumber \\
 & =\sum_{k,l,m,n}c_{k}c_{l}c_{m}c_{n}-6\sum c_{k}^{2}c_{l}c_{m}+\sum_{k,l}(8c_{k}^{3}c_{l}+3c_{k}^{2}c_{l}^{2})-6\sum_{k}c_{k}^{4}.\label{eq:cccc_final}
\end{align}
 Similarly, the formula for $I_{\textrm{{ssss}}}$ is obtained as

\begin{align}
I_{\textrm{{ssss}}} & =\sum_{k,l,m,n}s_{k}s_{l}s_{m}s_{n}-6\sum_{k,l,m}s_{k}^{2}s_{l}s_{m}+\sum_{k,l}(8s_{k}^{3}s_{l}+3s_{k}^{2}s_{m}^{2})-6\sum_{k}s_{k}^{4}.\label{eq:ssss_final}
\end{align}
Combining the results of the Eqs.~\eqref{eq:ccss3-1-1}, \eqref{eq:cccc_final},
and \eqref{eq:ssss_final}, we obtain 

\begin{eqnarray}
I_4= &  & I_{{\rm cccc}}+I_{{\rm ssss}}+2I_{{\rm ccss}}\nonumber \\
= &  & \sum_{k,l,m,n}c_{k}c_{l}c_{m}c_{n}+s_{k}s_{l}s_{m}s_{n}+2c_{k}c_{l}s_{m}s_{n}\nonumber \\
 &  & -\sum_{k,l,m}\left(6c_{k}^{2}c_{l}c_{m}+6s_{k}^{2}s_{l}s_{m}+2c_{k}^{2}s_{l}s_{m}+2c_{k}c_{l}s_{m}^{2}+8c_{k}s_{k}c_{l}s_{m}\right)\nonumber \\
 &  & +\sum_{k,l}\left(8c_{k}^{3}c_{l}+3c_{k}^{2}c_{l}^{2}+8s_{k}^{3}s_{l}+3s_{k}^{2}s_{m}^{2}+8c_{k}^{2}s_{k}s_{l}+2c_{k}^{2}s_{l}^{2}+8c_{k}c_{l}s_{l}^{2}+4c_{k}s_{k}c_{l}s_{l}\right)\nonumber \\
 &  & -6\sum_{k}(c_{k}^{4}+s_{k}^{4}+2c_{k}^{2}s_{k}^{2}).
\end{eqnarray}
This is equivalent to Eq.~\eqref{eq:statement1}.  \proofend

%- - - - - - - - - - - - - - - - - - - - - - - - - - - - - - - - - - - - - - - -%
\subsection{Proof of Eq.~(\ref{eq:statement2})}
\label{sec:Q2}

Using the continuous distribution formalism one can write $I_4$ from
Eq.~(\ref{eq:S}) as
\begin{equation}
	I_4=\frac{N!}{(N-4)!}\int {\rm d} \vec z_4 f^{\vec z_N^\indchainV}_4(\vec z_4) \Big(c_1 c_2 c_3 c_4+ s_1 s_2 s_3 s_4
  +2 c_1 c_2 s_3 s_4\Big),
  \label{eq:I4S}
\end{equation}
where $f^{\vec z_N^\indchainV}_4(\vec z_4)$ is the reduced 4-body correlation 
function for the chain, cf. \actualize{Eq.~(73)}{Eq.~(\ref{eq:I4av})} of the main text.
It is computed as
$f_4^{\vec z_N^\indchainV}(\vec z_4)=\int {\rm d}z_5\cdots {\rm d}z_N 
f^{\vec z_N^\indchainV}_N(\vec z_N)$ from the permutationally invariant
\( N \)-variate probability density of $N$ particles with the
$z$-coordinates  $\vec z_N^\indchainV$, which is given by
\begin{equation}\label{eq:fNz0S}
f_N^{\vec z_N^\indchainV}(\vec z_N)=\frac{1}{N!}\sum_{\pi\in S_N}\prod_{k=1}^N \delta\left(z_k-z_{\pi(k)}^\indchainV\right)
=\frac{1}{N!}\sum_{\neq(k_1,k_2,\ldots,k_N)}\prod_{j=1}^N \delta\left(z_j-z_{{k_j}}^\indchain\right)\,.
\end{equation}
Here, $S_N$ is the
permutation group of $N$ particles and the first sum runs over all permutations
$\pi$ from that group.
In the second sum 
the \( k_j \) indices ranges from 1 to \( N \) with the restriction that the indices be different, and
 \(z_k^\indchain\) are the locations of the particles on the chain. 

For {\it any} permutationally invariant probability density $f_N$ 
one can show that 
\begin{equation}\label{eq:I4e}
\int {\rm d} \vec z_4 f_4 \Big(c_1 c_2 c_3 c_4+ s_1 s_2 s_3 s_4
  +2 c_1 c_2 s_3 s_4\Big)=
\int\mathrm{d}\vec z_4 f_{4}\cos\Big[\frac{z_1-z_2+z_3-z_4}{L}\Theta\Big]
=\int\mathrm{d}\vec z_4 f_{4}e^{i\frac{z_1-z_2+z_3-z_4}{L}\Theta}
\end{equation}
holds. The second equality holds because the sine expression occuring in the
exponent is an odd function under the exchange of $z_1+z_3$ and $z_2+z_4$.
In this way, we can express $I_4$ with the help of characteristic functions.
In particular, the multivariate characteristic of the
multivariate probability density $f_N(\vec z_N)$ 
is
\begin{equation}
\label{eq:fNhat}
\hat{f}_N( {\vec \alpha}_N)=\left\langle e^{i\sum_{k=1}^N \alpha_k z_k}\right\rangle\
=\int {\rm d} \vec z_N  f_N(\vec z_N)e^{i\sum_{k=1}^N \alpha_k z_k}.
\end{equation} 
As can be easily checked, 
$\hat f_N({\vec \alpha}_N)$
has the following properties: 
(i) $\hat f_N$ is permutationally invariant if $f_N$ is, 
(ii) $\hat f_N\Big([\alpha_1,\alpha_2,...,\alpha_{N-1},0]\Big)=\hat f_{N-1}({\vec \alpha}_{N-1})$,
(iii) $\hat f_N({\vec 0}_N)=1$, where ${\vec 0}_N$ is a vector where all entries
are equal to $0$, and
(iv) $f_N(-\vec \alpha_N)=f_N(\vec \alpha_N)^*$.  
Property (ii) follows from the fact that 
$\int {\rm d} z_N f_N({\vec z}_N)=f_{N-1}({\vec z}_{N-1})$ 
and property (iii) follows from the
normalization of $f_N({\vec z}_N)$.

Comparing the Equations~(\ref{eq:I4}), (\ref{eq:I4e}), and (\ref{eq:fNhat}), we observe that 
\begin{equation}\label{eq:I4_vs_f4hat}
  I_4=\frac{N!}{(N-4)!}\hat f^{\vec z_N^\indchainV}_4(\alpha,-\alpha,\alpha,-\alpha),\ \ \mathrm{where}\ \ \alpha=\frac{\Theta}{L},
\end{equation}
and where $\hat f^{\vec z_N^\indchainV}_4$ has to be computed from the probability density
of Eq.~(\ref{eq:fNz0S}).
In general, the lower elements $f^{\vec z_N^\indchainV}_M(\vec z_M)$ 
(\( M\leq N \))  
are given by 
 \begin{equation}\label{eq:fMS}
f_M^{\vec z_N^\indchainV}(\vec z_M)=\frac{(N-M)!}{N!}\sum_{\neq(k_1,k_2,\ldots,k_M)}\prod_{j=1}^M \delta\left(z_j-z_{{k_j}}^\indchain\right)\,.
\end{equation}
Let us compute the characteristic function of $f_M^{\vec z_N^\indchainV}$, dropping from
now on 
the upper index ${\vec z_N^\indchainV}$ in order to simplify the notation. 
We obtain the following recurrence relation
\begin{eqnarray}
\hat{f}_M(\vec\alpha_M)&=&\frac{(N-M)!}{N!}\sum_{\neq(k_1,k_2,\ldots,k_M)}e^{i\sum_{j=1}^M \alpha_j z_{{k_j}}^0}\nonumber\\
&=&\frac{(N-M)!}{N!}\left\{\sum_{\neq(k_1,k_2,\ldots,k_{M-1}),k_M}e^{i\sum_{j=1}^M \alpha_j z_{{k_j}}^0}-\sum_{\neq(k_1,k_2,\ldots,k_{M-1})}\sum_{l=1}^{M-1}e^{i\sum_{j=1}^{M-1} \alpha_j z^0_{{k_j}}} e^{i \alpha_M z^0_{{k_l}}}\right\}\nonumber\\
&=&\frac{(N-M)!}{N!}\left\{N\left\langle e^{i \alpha_M z_1}\right\rangle \frac{N!}{(N-M+1)!}\left\langle e^{i\sum_{j=1}^{M-1} \alpha_j z^0_j}\right\rangle-\frac{N!}{(N-M+1)!}\sum_{l=1}^{M-1}\left\langle e^{i\sum_{j=1}^{M-1} \alpha_j z^0_j} e^{i\alpha_M z^0_l} \right\rangle\right\}\nonumber\\
&=&\frac{1}{N-M+1}\left\{N\hat{f}_1\left(\alpha_M\right)\hat{f}_{M-1}(\vec\alpha_{M-1})- \sum_{l=1}^{M-1}\hat{f}_{M-1}(\vec \alpha_{M-1}+\alpha_M\hat e_l)\right\},
\label{eq:recurrence}
\end{eqnarray}
where $\hat e_l$ is a vector of length $M-1$ that has only one
nonvanishing element (that is equal to 1) at the position $l$.
It can be used to compute $I_4$ {\it via} Eq.~(\ref{eq:I4_vs_f4hat}), leading to
\begin{equation}\label{eq:I4_rec1}
\hat{f}_4(\alpha,-\alpha,\alpha,-\alpha )= \frac{1}{N-3}\left\{N\hat{f}^*_1(\alpha)\hat{f}_3(\alpha,-\alpha,\alpha)-
2\hat{f}_2(\alpha,-\alpha)+\hat{f}_3(\alpha,-2\alpha,\alpha)\right\}
\end{equation}
We can apply the recurrence relation again for $M=3$ in order to reduce the complexity
of this expression. We obtain
\begin{eqnarray}
\hat{f}_3(\vec \alpha_3)&=&
\frac{1}{(N-1)(N-2)}\left\{N^2\hat{f}_1(\alpha_1)\hat{f}_1(\alpha_2)\hat{f}_1(\alpha_3)
+2\hat{f}_1(\alpha_1+\alpha_2+\alpha_3)\right.\nonumber\\
&&
\left. N\big[\hat{f}_1(\alpha_1)\hat{f}_1(\alpha_2+\alpha_3)+\hat{f}_1(\alpha_2)\hat{f}_1(\alpha_1+\alpha_3)+
\hat{f}_1(\alpha_3)\hat{f}_1(\alpha_1+\alpha_2)\big]\right\}\,,\nonumber
\end{eqnarray}
which for the two cases of interest in Eq.~(\ref{eq:I4_rec1}) yields
\begin{eqnarray}
\hat{f}_3(\alpha,-\alpha,\alpha)&=&\frac{1}{(N-1)(N-2)}\left\{N^2\left|\hat{f}_1(\alpha)\right|^2 \hat{f}_1(\alpha)-N\hat{f}_1(2\alpha)\hat{f}_1^*(\alpha) -2(N-1)\hat{f}_1(\alpha)\right\}
\,,\nonumber\\
\hat{f}_3(\alpha,-2\alpha,\alpha)&=&\frac{1}{(N-1)(N-2)}\left\{N^2\hat{f}_1^2(\alpha)\hat{f}_1^*(2\alpha)-2N\left|\hat{f}_1(\alpha)\right|^2-N\left|\hat{f}_1(2\alpha)\right|^2+2\right\}\,.\nonumber
\end{eqnarray}
Similarly, we obtain for \( M=2 \) that
 \begin{equation}
\hat{f}_2\left(\alpha_1,\alpha_2\right)= \frac{1}{N-1}\left\{N\hat{f}_1\left(\alpha_1\right)\hat{f}_1\left(\alpha_2\right)-\hat{f}_1\left(\alpha_1+\alpha_2\right)\right\}\, 
\end{equation}
For the special case of interest
 \( \alpha_1=-\alpha_2 \) occuring in Eq.~(\ref{eq:I4_rec1}) this reduces to
\begin{equation}
\hat{f}_2\left(\alpha,-\alpha\right)=\frac{1}{N-1}\left\{N\left|\hat{f}_1\left(\alpha\right)\right|^2-1\right\}\,.
\end{equation}
%%XXXX%%%
Finally
\begin{eqnarray*}
\hat{f}_4(\alpha,-\alpha,\alpha,-\alpha )&=& \frac{1}{(N-1)(N-2)(N-3)}\left\{2(N-3)-4 N(N-2)\left|\hat{f}_1(\alpha)\right|^2+N^3\left|\hat{f}_1(\alpha)\right|^4 + 
N\left|\hat{f}_1(2\alpha)\right|^2\right.\nonumber\\
&&\qquad\qquad\qquad\qquad\qquad\qquad\left.-2N^2\textrm{Re}\big[\hat{f}_1^2(\alpha)\hat{f}_1(2 \alpha)^*\big]\right\},
\end{eqnarray*}
which due to the identity~(\ref{eq:I4_vs_f4hat}) is equivalent to Eq.~(\ref{eq:statement2}) for
$\hat f_1(\alpha)=\frac{1}{N}\sum_{k}e^{i\alpha z_k^\indchain}$ [computed with 
the Eqs.~(\ref{eq:fNhat}) and (\ref{eq:fMS})] with $\alpha=\frac{\Theta}{L}$.
\proofend
%Inserting these results in the expression for \( J_x^4(\Theta/L)\), we obtain
%\begin{eqnarray}
%\left\langle  J_x^4\left(\frac{\Theta}{L}\right)\right\rangle&=&\frac{N^2}{16(N-1)(N-3)}\left\{(3N+1)(N-3)-2(3N-7)N \left|\hat{f}_1\left(\frac{\Theta}{L}\right)\right|^2 +3 N^2 \left|\hat{f}_1\left(\frac{\Theta}{L}\right)\right|^4+3\left|\hat{f}_1\left(\frac{2\Theta}{L}\right)\right|^2\right.\nonumber\\
%&&\qquad\qquad\qquad\qquad\qquad\left.-3N\left[\left(\hat{f}_1\left(\frac{\Theta}{L}\right)^*\right)^2\hat{f}_1\left(\frac{2\Theta}{L}\right)+\left(\hat{f}_1\left(\frac{\Theta}{L}\right)\right)^2\hat{f}_1\left(\frac{2\Theta}{L}\right)^*\right]\right\}\nonumber
%\end{eqnarray}

%----------------------------------------------------------------------------%
\section{Additional calculations for obtaining $(\Delta\Theta)_{\rm s}^{-2}|_{\Theta=0}$
for Observation \actualize{7}{7}}

We will show that for $\Theta\to 0$, the inverse variance 
of the estimation of $\Theta$ is given by

\begin{equation}
\left(\Delta\Theta\right)_{\rm s}^{-2}
|_{\Theta=0}=\frac{N}{L^{2}}\left[\sigma^{2}-\mathrm{cov}(z_{1},z_{2})\right],\label{eq:deltatheta2}
\end{equation}
 where 
\begin{align}
\sigma^{2} & =\int{\rm d}z_1 f_{1}(z_1)(z_1-\langle z_1\rangle)^{2},\nonumber \\
\langle z_1\rangle & =\int{\rm d}z_1 f_{1}(z_1)z_1, \nonumber\\
\mathrm{cov}(z_1,z_2) & = \int {\rm d}z_1 {\rm d}z_2 f_2(z_1,z_2)
 (z_1-\langle z_1\rangle)(z_2-\langle z_2\rangle)=\langle z_1 z_2\rangle
  -\langle z_1\rangle \langle z_2\rangle.
\end{align}
\emph{Proof.} We estimate the uncertainty from the error propagation formula
\begin{equation}
\label{eq:DTheta}
(\Delta\Theta)_{\rm s}^{2}=\frac{(\Delta  J_{x}^{2})_{\rm s}}{|\partial_{\Theta}\langle  J_{x}^{2}\rangle_{\rm s}|^{2}},
\end{equation}
cf. \actualize{Eq.~(49)}{Eq.~(\ref{DeltaTheta})} of the main text, for general continuous density
profiles. The quantities which occur are 
$\langle J_x^2\rangle_{\rm s}(\Theta)$,
$\partial_\Theta \langle J_x^2\rangle_{\rm s}(\Theta)$ and
$\langle J_x^4\rangle_{\rm s}(\Theta)$.
In order to get the 
desired limit, we need to expand them around $\Theta=0$. Let us start with 
$\langle J_x^2\rangle_{\rm s}(\Theta)$. For fixed particle positions 
$\vec z_N$, we obtain
\begin{align*}
\langle  J_{x}^{2}\rangle_{\rm s}^{\vec z_N}(\Theta) & =\frac{N\hbar^2}{4}
\Big[1-\frac{1}{N(N-1)}\sum_{n\neq m}\cos\Big(\frac{z_n-z_m}{L}\Theta\Big)
\Big],
\end{align*}
from the Eqs.~(\actualize{38,39}{\ref{eq:I2},\ref{Jx2c}}) and (\actualize{44}{\ref{ccss_cos}}).
Averaging this over a general permutationally
independent density profile $f_N(\vec z_N)$, we obtain
\begin{align*}
\langle  J_{x}^{2}\rangle_{\rm s}^{f_N}(\Theta) & =
\frac{N\hbar^2}{4}\left[
1-\int{\rm d}z_1 {\rm d}z_2 f_2(z_1,z_2)\cos\Big(\frac{z_1-z_2}{L}\Theta\Big)
\right]\equiv
\frac{N\hbar^2}{4}(1-\tilde I_2).
\end{align*}
Expanding the cosine in the integral we arrive at
\begin{align}
\tilde I_2 & \approx 1-\frac{1}{2L^2}\int{\rm d}z_1 {\rm d}z_2 f_2(z_1,z_2) (z_1-z_2)^2\ \Theta^2
+O(\Theta^4)\nonumber\\
&
=
1 - \frac{1}{L^2} \Big(\int{\rm d}z_1 f_1(z_1) z_1^2 -
   \int{\rm d}z_1 {\rm d}z_2 f_2(z_1,z_2) z_1 z_2\Big)\Theta^2+O(\Theta^4)\nonumber\\
&
=1-\frac{1}{L^2}\Big(\sigma^2-\mathrm{cov}(z_1,z_2)\Big)\Theta^2+O(\Theta^4),
\label{eq:I2exp}
\end{align}
where the last line is obtained by adding and subtracting a term 
$\langle z_1\rangle^2$. We also used that due to the permutational invariance
$\langle z_1^2\rangle=\langle z_2^2\rangle$ and $\langle z_1\rangle=\langle z_2\rangle$ hold. 
This leads to 
\begin{align}
\langle  J_{x}^{2}\rangle_{\rm s}^{f_N}(\Theta) & 
\approx \frac{N\hbar^2}{4L^2}\Big(\sigma^2-\mathrm{cov}(z_1,z_2)\Big)\Theta^2+O(\Theta^4)
\label{eq:Jx2exp}
\end{align}
and
\begin{equation}
\partial_\Theta \langle  J_{x}^{2}\rangle_{\rm s}^{f_N}(\Theta)
 \approx \frac{N\hbar^2}{2L^2}\Big(\sigma^2-\mathrm{cov}(z_1,z_2)\Big)\Theta+O(\Theta^3),
 \label{eq:dJx2exp}
\end{equation}
Let us now consider the expansion of the term $\langle J_x^4\rangle_{\rm s}(\Theta)$. Again for fixed positions $\vec z_N$, we have 
[cf. \actualize{Eq.~(68)}{Eq.~(\ref{Jx4fff})}]
\begin{equation*}
\frac{\langle  J_{x}^4\rangle_{\rm s}^{\vec z_N}(\Theta)}{\hbar^4}  
=\frac{3N^2-2N}{16}-\frac{3N-4}{8(N-1)}\sum_{k\neq l}
\cos\Big(\frac{z_n-z_m}{L}\Theta\Big) + \frac{3}{16}\frac{1}{(N-1)(N-3)} I_4,
\end{equation*}
with $I_4$ from Eq.~(\ref{eq:S}) above. 
Note that in contrast to this
equation, the particle positions are labelled by $\vec z_N$ instead
of $\vec z_N^\indchainV$ because we have to average the expression over $f_N$.
This leads to
\begin{align}
\frac{\langle  J_{x}^{4}\rangle_{\rm s}^{f_N}(\Theta)}{\hbar^{4}} & =\frac{3N^2-2N}{16}-\frac{N(3N-4)}{8}\int\mathrm{d}z_{1}\mathrm{d}z_{2}\, f_{2}(z_{1},z_{2})\,\cos\Big(\frac{z_{1}-z_{2}}{L}\Theta\Big)\nonumber \\
 & +\frac{3N(N-2)}{16}\int\mathrm{d}^{4}z f_{4}(\vec z_4)
 \Big(c_1 c_2 c_3 c_4+s_1 s_2 s_3 s_4
+2 c_1 c_2 s_3 s_4\Big)\nonumber
\\
&
\equiv
\frac{3N^2-2N}{16}-\frac{N(3N-4)}{8}\tilde I_2
 +\frac{3N(N-2)}{16}\tilde I_4,\label{eq:Jx4}
\end{align}
where we used again the permutational invariance of $f_N$. We need to expand
the expression $\tilde I_4$. Using the first equality from Eq.~(\ref{eq:I4e})
and expanding the occurring cosine as before one obtains that 
\begin{align}
\tilde I_4 
&
\approx 
1-\frac{1}{2L^2}\int\mathrm{d}^{4}z f_{4}(z_{1},z_{2},z_{3},z_{4})
 (z_1+z_2-z_3-z_4)^2\Theta^2+O(\Theta^4)\nonumber\\
 &
 =1-\frac{2}{L^2}[\sigma^2-\mathrm{cov}(z_1,z_2)]\Theta^2+O(\Theta^4).
 \label{eq:I4exp}
\end{align}
Inserting the expansions of $\tilde I_2$ from Eq.~(\ref{eq:I2exp})
and of $\tilde I_4$ from Eq.~(\ref{eq:I4exp}) into Eq.~(\ref{eq:Jx4}) leads to
\begin{align}
\frac{\langle  J_{x}^{4}\rangle_{\rm s}^{f_N}(\Theta)}{\hbar^{4}} & 
\approx
\frac{3N^2-2N}{16}
-\frac{N(3N-4)}{8}
\Big(1-\frac{1}{L^2}[\sigma^2-\mathrm{cov}(z_1,z_2)]\Theta^2\Big)
\nonumber\\
&+\frac{3N(N-2)}{16}
\Big(1-\frac{2}{L^2}[\sigma^2-\mathrm{cov}(z_1,z_2)]\Theta^2\Big)
+O(\Theta^4)\nonumber\\
&
= \frac{N}{4L^2}[\sigma^2-\mathrm{cov}(z_1,z_2)]\Theta^2+O(\Theta^4).
\label{eq:Jx4exp}
\end{align}
Now we have all the necessary ingredients to prove the claim. Indeed,
inserting the Eqs.~(\ref{eq:Jx2exp},\ref{eq:dJx2exp},\ref{eq:Jx4exp}) 
into Eq.~(\ref{eq:DTheta}) we obtain
\begin{equation*}
	(\Delta\Theta)_{\rm s}^{-2} \approx \frac{\frac{N^2\hbar^4}{4L^4}[\sigma^2-\mathrm{cov}(z_1,z_2)]^2\Theta^2+O(\Theta^4)}{\frac{N\hbar^4}{4L^2}[\sigma^2-\mathrm{cov}(z_1,z_2)]\Theta^2+O(\Theta^4)}
=
\frac{N}{L^2}[\sigma^2-\mathrm{cov}(z_1,z_2)]+O(\Theta^2),
\end{equation*}
which proves Eq.~(\ref{eq:deltatheta2}). \proofend

%----------------------------------------------------------------------------%


\begin{thebibliography}{99}

\bibitem{Hald1999}J. Hald, J. L. S\o rensen, C. Schori, and E. S. Polzik, Phys. Rev. Lett. {\bf83}, 1319 (1999).
\bibitem{Julsgaard2001}B. Julsgaard, A. Kozhekin and E. S. Polzik, Nature {\bf 413}, 400 (2001).

\bibitem{Meyer2001}V. Meyer, M. A. Rowe, D. Kielpinski, C. A. Sackett, W. M. Itano, C. Monroe, and D. J. Wineland, Phys. Rev. Lett. {\bf 86}, 5870 (2001).

\bibitem{Hammerer2010}K. Hammerer, A. S. S\o rensen, and E. S. Polzik, Rev. Mod. Phys. {\bf82}, 1041 (2010).

%QND
\bibitem{Duan2000} L. M. Duan, J. I. Cirac, P. Zoller, and E. S. Polzik, Phys. Rev. Lett.  {\bf85}, 5643 (2000). 
\bibitem{Kuzmich1998}A. Kuzmich, N. P. Bigelow, and L. Mandel, Europhys. Lett. {\bf 42}, 481 (1998).

% Quantum teleportation between light and matter
\bibitem{Sherson2006} J.F. Sherson, H. Krauter, R.K. Olsson, B. Julsgaard, K. Hammerer, I. Cirac, and E.S. Polzik, Nature (London) {\bf 443}, 557 (2006).

% Magnetometry with cold atomic ensembles
% Differential magnetometry with two ensembles
\bibitem{Wasilewski2010}W. Wasilewski, K. Jensen, H. Krauter, J.J. Renema, M.V. Balabas, and E.S. Polzik, Phys. Rev. Lett. {\bf104}, 133601 (2010).
\bibitem{Eckert2006} K. Eckert, P. Hyllus, D. Bru\ss, U.V. Poulsen, M. Lewenstein, C. Jentsch, T. Muller, E.M. Rasel, and W. Ertmer, Phys. Rev. A {\bf 73}, 013814 (2006).
% Squeezed-Light Optical Magnetometry
\bibitem{Wolfgramm2010} F. Wolfgramm, A. Cer\`e, F. A. Beduini,
A. Predojevi\'c, M. Koschorreck, and M.W. Mitchell,  Phys. Rev. Lett. {\bf 105}, 053601 (2010).
% Spin-squeezing of a large-spin system via QND measurement
\bibitem{Sewell2011} R.J. Sewell, M. Koschorreck, M. Napolitano, B. Dubost, N. Behbood, and M.W. Mitchell, hys. Rev. Lett. {\bf 109}, 253605 (2012).
% High Bandwidth Atomic Magnetometery with Continuous Quantum Nondemolition Measurements
\bibitem{ShahPRL2010} V. Shah, G. Vasilakis, and M.V. Romalis,
Phys. Rev. Lett. {\bf 104}, 013601 (2010).
% Quantum-enhanced magnetometer with low-frequency squeezing
\bibitem{HorromPRA2012}
T. Horrom, R. Singh, J.P. Dowling, and E.E. Mikhailov,
Phys. Rev. A {\bf 86}, 023803 (2012).

% CV description of spin ensembles
\bibitem{Giedke2002}G. Giedke and J.I. Cirac, Phys. Rev. A {\bf66}, 032316 (2002).
\bibitem{Madsen2004}L. B. Madsen and K. M\o lmer, Phys. Rev. A {\bf70}, 052324 (2004).
\bibitem{Hammerer2004} K. Hammerer, K. M\o lmer, E.S. Polzik, and J.I. Cirac, Phys. Rev. A {\bf70}, 044304 (2004).
 \bibitem{deEchaniz2005} S.R. de Echaniz, M.W. Mitchell, M. Kubasik, M. Koschorreck, H. Crepaz, J. Eschner, and E.S. Polzik, J. Opt. B {\bf 7}, S54 (2005).
\bibitem{Koschorreck2009} M. Koschorreck and M.W. Mitchell, J. Phys. B: At. Mol. Opt. Phys. {\bf42} 19550 (2009).

% Spin squeezing
\bibitem{Kitagawa1993}M. Kitagawa and M. Ueda, Phys. Rev. A {\bf47}, 5138 (1993). 
\bibitem{Wineland1994}D.J. Wineland, J.J. Bollinger, W. M. Itano, and D. J. Heinzen, Phys. Rev. A {\bf50}, 67 (1994). 
\bibitem{Sorensen2001}A.S. S\o rensen, L.M. Duan, J.I. Cirac, and P. Zoller, Nature (London) {\bf409}, 63 (2001).
\bibitem{Ma2011}J. Ma, X. Wang, C.P. Sun, and F. Nori, Phys. Rep. {\bf509}, 89 (2011).

% Compensated high temperature SQUID gradiometer for mobile NDE in magnetically noisy environments
\bibitem{Keenan2012} S.T. Keenan and E.J. Romans, NDT \&  E International {\bf 47}, 1 (2012).

% Bipartite singlets
\bibitem{Cable2010} H. Cable and G.A. Durkin, Phys. Rev. Lett. {\bf105}, 013603 (2010).
% Multipartite singlet by spin squeezing
\bibitem{Toth2010} G. T\'oth and M.W. Mitchell, New J. Phys. {\bf 12}, 053007 (2010).
\bibitem{Behbood2012} N. Behbood, M. Napolitano, G. Colangelo, B. Dubost, S. Palacios \'Alvarez, R. Sewell, G. T\'oth, and M. Mitchell, {\it  Generation of a Macroscopic Singlet State in an Atomic Ensemble} in Research in Optical Sciences, OSA Technical Digest No. QM1B.2 (Optical Society of America, Washington, D.C., 2012).

% Many-body singlets by dynamic spin polarization
\bibitem{Yao2011} W. Yao, Phys. Rev. B {\bf 83}, 201308(R) (2011).
% Engineering steady states using jump-based feedback for multipartite entanglement generation (singlet state generation)
\bibitem{Stevnson2011} R.N. Stevenson, J.J. Hope, A.R.R. Carvalho, Phys. Rev. A {\bf 84}, 022332 (2011).

% Interferometric measurement of local spin fluctuations in a quantum gas
\bibitem{Meineke2012} J. Meineke, J.-P. Brantut, D. Stadler, T. M\"uller, H. Moritz, and T. Esslinger, Nature Phys. {\bf 8}, 455 (2012).

% Experiments and theoretical proposals to create singlets
\bibitem{Singlet1}  K. Eckert, O. Romero-Isart, M. Rodriguez, M. Lewenstein, E.S. Polzik, and A. Sanpera, Nature Phys. {\bf 4}, 50 (2008).
\bibitem{Singlet2} T. Keilmann and J.J. Garc\'{\i}a-Ripoll, Phys. Rev. Lett.  {\bf 100}, 110406 (2008).
\bibitem{Singlet3} K. Ueda, H. Kontani, M. Sigrist, and P.A. Lee, Phys. Rev. Lett. {\bf 76}, 1932 (1996).
\bibitem{Singlet4} S. Miyahara and K. Ueda, Phys. Rev. Lett. {\bf 82}, 3701 (1999).
% Adiabatic Preparation of a Heisenberg Antiferromagnet Using an Optical Superlattice
\bibitem{Singlet5} M. Lubasch, V.  Murg, U. Schneider, J.I. Cirac, and M.-C. Ba\~nuls, Phys. Rev. Lett. {\bf 107}, 165301 (2011).

% Precisely mapping the magnetic field gradient in vacuum with an atom interferometer (experiment)
\bibitem{Zhou2010} M.-K. Zhou, Z.-K. Hu, X.-C. Duan, B.-L. Sun, J.-B. Zhao, and J. Luo,
Phys. Rev. A {\bf 82}, 061602(R) (2010).
% Sensing electric and magnetic fields with Bose-Einstein Condensates
% "magnetic microscopes"
\bibitem{Wildermuth2006} S. Wildermuth, S. Hofferberth, I. Lesanovsky, S. Groth, P. Kr\"uger, and J. Schmiedmayer, Appl. Phys. Lett. {\bf88}, 264103 (2006).
% High-Resolution Magnetometry with a Spinor Bose-Einstein Condensates
\bibitem{Vengalattore2007}  M. Vengalattore, J. M. Higbie, S. R. Leslie, J. Guzman, L. E. Sadler, and D. M. Stamper-Kurn,  Phys. Rev. Lett. {\bf98}, 200801 (2007).
% High resolution magnetic vector-field imaging with cold atomic ensembles
\bibitem{Koschorreckt2011} M. Koschorreck, M. Napolitano, B. Dubost, and M.W. Mitchell, Appl. Phys. Lett. {\bf 98}, 074101 (2011).
% Real-time vector field tracking with a cold-atom magnetometer
\bibitem{BM13} N. Behbood, F. Martin Ciurana, G. Colangelo, M. Napolitano, M. W. Mitchell, R. J. Sewell, 
Appl. Phys. Lett. {\bf 102}, 173504 (2013).

% Ensembles of spin-j particles:
% Efficient quantification of non-Gaussian spin distributions
\bibitem{Dubost2011} B. Dubost, M. Koschorreck, M. Napolitano, N. Behbood, R.J. Sewell, and M.W. Mitchell, Phys. Rev. Lett. {\bf 108}, 183602 (2012).
% Metrological experiment with a Dicke state
\bibitem{LS12} B. L\"ucke, M. Scherer, J. Kruse, L. Pezz\'e, F. Deuretzbacher, P. Hyllus, O. Topic, J. Peise, W. Ertmer, J. Arlt, L. Santos, A. Smerzi, and C. Klempt, Science {\bf 334}, 773 (2011).
% Spin-Nematic Squeezed Vacuum in a Quantum Gas: Supplementary Information
\bibitem{Hamley2012}  C.D. Hamley, C.S. Gerving, T.M. Hoang, E.M. Bookjans and M.S. Chapman,  Nature Phys. {\bf 8}, 305 (2012).
% Qutrit squeezing via semiclassical evolution
\bibitem{Klimov2011} A.B. Klimov, H. Tavakoli Dinani, Z.E.D. Medendorp, and H. de Guise, New J. Phys. {\bf 13}, 113033 (2011).

%  Spin squeezing of high-spin, spatially extended quantum fields.
\bibitem{Sau2011}  J.D. Sau, S.R. Leslie, M.L. Cohen, and D.M. Stamper-Kurn, New J. Phys. {\bf 12}, 085011 (2010).
% Quantum noise, scaling, and domain formation in a spinor Bose-Einstein condensate.
\bibitem{Mias2008} G.I. Mias, N.R. Cooper, and S.M. Girvin, Phys. Rev. A {\bf 77}, 023616 (2008).
% Multilevel Holstein-Primakoff approximation and its application to atomic spin squeezing and ensemble quantum memories
\bibitem{Kurucz2011} Z. Kurucz and K. M{\o}lmer, Phys. Rev. A {\bf 81}, 032314 (2010).
% Spinor Bose gases: Explorations of symmetries, magnetism and quantum dynamics
\bibitem{StamperKurn2012} D.M. Stamper-Kurn and M. Ueda, arXiv:1205.1888.

% Entanglement and Extreme Spin Squeezing
\bibitem{Sorensen2001b} A. S\o rensen and  K. M\o lmer, Phys. Rev. Lett. {\bf 86}, 4431 (2001).
\bibitem{Duan2002} L.-M. Duan, J.I. Cirac, and P. Zoller, \pra {\bf 65}, 033619 (2002).
\bibitem{Mustecaplioglu2002} \"O.E. M\"ustecapl{\i}o\u{g}lu, M. Zhang, and L. You,
Phys. Rev. A {\bf 66}, 033611 (2002),
\bibitem{Toth2004} G. T\'oth, Phys. Rev. A {\bf 69}, 052327 (2004).
\bibitem{Wiesniak2005} M. Wie{\'s}niak, V. Vedral, and  \v{C}. Brukner, New J. Phys. {\bf 7}, 258 (2005).
\bibitem{Vitagliano2011} G. Vitagliano, P. Hyllus, I.L. Egusquiza, and G. T\'oth, Phys. Rev. Lett. {\bf 107}, 240502 (2011).
% Planar quantum squeezing and atom interferometry
\bibitem{He2011} Q. Y. He, S.-G. Peng, P. D. Drummond, and M. D. Reid, Phys. Rev. A {\bf 84}, 022107 (2011).

\bibitem{Cirac1999} J.I. Cirac, A. K. Ekert, and C. Macchiavello, Phys. Rev. Lett. {\bf82},  4344 (1999).

\bibitem{TothPRL} G. T\'oth, W. Wieczorek, D. Gross, R. Krischek, C. Schwemmer, H. Weinfurter, Phys. Rev. Lett. {\bf105}, 250403 (2010). 

\bibitem{JC12}
Alternatively, one can argue based on the group theory of angular momentum.
Let us write the permutationally invariant density matrix in the usual block diagonal form.
Here, SU(2) operations act within the blocks, while permutations act between the blocks.
For the permutationally invariant case, all blocks corresponding to the same $j$ and $j_z$
are the same.  The block size corresponding to
$j=0$ is $1\times 1. $Thus, all permutationally invariant states can be written as 
%\bea
$\varrho=(1-p_{\rm s})\varrho_{>0}+p_{\rm s}\varrho_{\rm s},\nonumber$
%\eea
where $0 \le p_{\rm s}\le 1,$ $\varrho_{\rm s}$ is defined in Eq.~(\ref{singlet_pairs}) and $\varrho_{>0}$ is a permutaionally invariant state not having a contribution in the $J^2=0$ subspace. 
See a very clear presentation of the group theoretical basis for this statement in another context in R.B.A. Adamson, P.S. Turner, M.W. Mitchell, and A.M. Steinberg, Phys. Rev. A {\bf 78}, 033832 (2008).
This comment is based on J. Calsamiglia, private communication (2012).


\bibitem{Note2}  Note that it has been known that a multiparticle singlet is within the space of the tensor products of two-particle singlets for $j=\frac{1}{2}$ \cite{Cirac1999}.
Here we have shown that the state Eq.~(\ref{singlet_J0}) is an {\it equal} mixture of such states.
One can see even by simple numerics for small systems that an analogous condition does not hold for $j>\frac{1}{2}.$

\bibitem{Note} Note that while there are $N!$ different permutation operators, there are only $(N+1)!!=(N+1)(N-1)(N-3)...$ different permutations of the tensor product of two-particle singlets. This can be proved as follows. 
We call $f(N)$ the number of different pairings for $N$ particles.
First, pair the first and the second particle. We will still have $f(N-2)$ possibilities of pairing all the remaining particles among each other.
So, we have $f(N)=f(N-2)+\text{other contributions}$. Second, we pair the first particle with the third one. Again we will have $f(N-2)$ possibilities of 
pairing the remaining spins. So, $f(N)=2f(N-2)+\text{other contributions}$. Then we pair the first particle with the fourth one, and again $f(N-2)$ particles.
We can repeat this procedure until we have paired the first particle with all of the others in the chain. As there are $N-1$ particles that are not the first particle, 
in the end we will have $f(N)=(N-1)f(N-2)$. We also know that $f(2)=1$ since for two particles there is only one possible pairing.
Solving this recurrence relation, we get $f(N)=(N-1)!!$. Note that we are careful and do not double count the permutations.

\bibitem{GreinerNat02}
M. Greiner, O. Mandel, T. Esslinger, T.W. H\"ansch, and I. Bloch,
Nature (London) {\bf 415}, 39 (2002).
%{\it Quantum phase transition from a superfluid to a Mott insulator in a gas 
%of ultracold atoms}

\bibitem{MorschRMP06}
O. Morsch and M. Oberthaler, Rev. Mod. Phys. {\bf 78}, 179 (2006).
%{\it Dynamics of Bose-Einstein condensates in optical lattices}

\bibitem{EPAPS} See Supplemental Material for additional derivations.

\bibitem{Note_SU2invariance}
{The fact that $\ex{j_k\otimes j_l}_{\varrho_{12}^{{\rm red},j}}=0$ for $k\neq l$
follows from the SU(2) invariance of $\varrho_{12}^{{\rm red},j}$ because
a rotation around $k$ by $\pi$ leads to $j_k\to j_k$ but $j_l\to -j_l$, 
whence $\ex{j_k\otimes j_l}_{\varrho_{12}^{{\rm red},j}}=-\ex{j_k\otimes j_l}_{\varrho_{12}^{{\rm red},j}}$.
In analogy, it follows that $\langle j_x^{(1)}j_y^{(2)}j_y^{(3)}j_y^{(4)}\rangle_{\rho_{1234}^{\rm red}}=
\langle j_x^{(1)}j_x^{(2)}j_x^{(3)}j_y^{(4)}\rangle_{\rho_{1234}^{\rm red}}=0$ as used in Sec.~\ref{sec:CalcPrec}.
}

\bibitem{noise} Noise can have a strong impact on quantum metrological setups. See 
R. Demkowicz-Dobrza\'nski, J. Ko\l ody\'nski, and M. Gu\unichar{539}\unichar{259}, Nat. Comm. {\bf 3}, 1063 (2012). 

\bibitem{SUdsinglet} Note that there is still a unique permutationally invariant $SU(d)$ singlet for a 
system of $d$-dimensional particles.

\bibitem{Gaussian} For a Gaussian probability  distribution, the first- and second-order moments determine all higher-order moment. Thus, if the outcome statistics of the $\Jc$ measurement is Gaussian, the fourth moment $\ex{\Jc^4}$ can be obtained from the second moment $\ex{\Jc^2}$ as 
$\ex{\Jc^4}=3\ex{\Jc^2}.$  

\bibitem{Fisher} I. Apellaniz and P. Hyllus, unpublished (2012).

\end{thebibliography}
\end{document}